\documentclass{phostproc}

\usepackage{overpic}
\usepackage{upgreek}
\usepackage{xcolor}
\newcommand{\blu}{\textcolor{blue} }

\title{Pulsating stars in binary systems: a review}
\author{Simon J. Murphy,$^{1,\dagger}$}

\affiliation{
$^{1}$ Sydney Institute for Astronomy (SIfA), School of Physics, University of Sydney, NSW 2006, Australia.\\
$^{\dagger}$ email: simon.murphy@sydney.edu.au\\
}

\shorttitle{Pulsating stars in binaries}
\shortauthors{Simon J. Murphy}

\abs{Binary systems anchor many of the fundamental relations relied upon in asteroseismology. Masses and radii are rarely constrained better than when measured via orbital dynamics and eclipse depths. Pulsating binaries have much to offer. They are clocks, moving in space, that encode orbital motion in the Doppler-shifted pulsation frequencies. They offer twice the opportunity to obtain an asteroseismic age, which is then applicable to both stars. They enable comparative asteroseismology -- the study of two stars by their pulsation properties, whose only fundamental differences are the mass and rotation rates with which they were born. In eccentric binaries, oscillations can be excited tidally, informing our knowledge of tidal dissipation and resonant frequency locking. Eclipsing binaries offer benchmarks against which the asteroseismic scaling relations can be tested. We review these themes in light of both observational and theoretical developments recently made possible by space-based photometry.}

\begin{document}

\maketitle


\section{Introduction}

Pulsating stars can now be found over almost the entire HR diagram \citep{aertsetal2010}, with new classes still being discovered \citep{pietrukowiczetal2017}. The field of asteroseismology has bloomed with the delivery of continuous and ultra-precise light curves from space telescopes such as {\it CoRoT} \citep{baglinetal2006} and \textit{Kepler} \citep{kochetal2010,boruckietal2010}, and is set to continue with the Transiting Exoplanet Survey Satellite ({\it TESS}; \citealt{rickeretal2015}) which has announced its first exoplanet discovery \citep{huangetal2019}. Although designed to detect transiting exoplanets, the photometry from these missions is ideal for asteroseismology, which plays a critical reciprocal role in providing stellar radii for better exoplanet characterisation.

Like planets and pulsators, binary stars are also ubiquitous, with most stars being members of binary or multiple systems \citep{duchene&kraus2013,moe&distefano2017,guszejnovetal2017}. They can be found throughout the HR diagram, with more massive stars having greater multiplicity. Double-lined spectroscopic binary systems (SB2s) have been indispensable to astronomy as a means of measuring dynamical masses of stars for over a hundred years \citep{stebbins1911}. The strong synergy between pulsating and binary stars in providing stellar radii and masses suggests that good astrophysical benchmarks exist at the interface.

Stellar spectra are not strictly necessary to measure the orbits of pulsating binary stars. In recent years, precise orbital elements have been derived by pulsation timing \citep{teltingetal2012,shibahashi&kurtz2012,murphyetal2014,murphyetal2018}, building upon the classical O-C (observed minus calculated) methods that were employed on ground-based data \citep{barnes&moffett1975,sterken2005c}. The quantifiable orbital parameters mirror those of the radial velocity (RV) method: these are the period, eccentricity, projected semi-major axis (i.e.\ $a \sin i$), longitude of periastron and the binary mass function. In fact, radial velocity is the time-derivative of the light arrival-time delay, $\tau$
\begin{eqnarray}
RV(t) = {\rm c} \frac{\rm{d} \tau}{{\rm d}t},
\end{eqnarray}
where c is the speed of light, so while the RV method has a sensitivity that scales as $P^{-1/3}$, pulsation timing goes as $P^{+2/3}$ (see footnote\footnote{In an earlier version of this article, there was a minus sign in equation 1. That mistake is rectified in this version, which now follows the convention established in footnote 2 of \citet{murphy&shibahashi2015}.}). The methods are therefore highly complementary, and data from each can be combined to deduce orbits to exquisite precision \citep{murphyetal2016b}.

A great advantage of modelling stars in binaries is that they can be assumed to share the same evolutionary history. Comparing two similar but unrelated stars can be difficult because of uncertainties in the relative age and metallicity of the objects, but for binaries this is not the case. Their pulsation properties can then be compared primarily as a function of mass, subject to differences in rotation rates. Any differences in chemical composition between main-sequence stars in a binary can be understood as the result of mass- and rotation-dependent mixing and diffusive processes \citep{charbonneau&michaud1991,turcotteetal2000}, unencumbered by the large star-to-star abundance variations that are common among isolated intermediate-mass stars \citep{hill&landstreet1993}.

Stellar inclinations are rarely determined better than for binary or pulsating stars (e.g.\ \citealt{thompsonetal2012}). Under the assumption of equipartition of energy, the inclination of a pulsating star can be judged from the amplitudes of rotationally-split non-radial modes \citep{gizon&solanki2003}. This typically only applies to stochastic oscillations, but slowly-rotating gamma Doradus ($\gamma$\,Dor) stars also follow suit \citep{kurtzetal2014,saioetal2015}. Although inclinations can rarely be determined from delta Scuti ($\delta$\,Sct) pulsations, knowledge of the inclination by other means can aid mode identification in these stars, as we shall see in Sect.\,\ref{sec:dsct}. In eclipsing binaries, the orbital inclinations can be determined precisely, and even ellipsoidal variability or well-constrained masses can offer some constraints. The extent to which binaries may suffer spin--orbit misalignment remains an active area of research \citep{albrechtetal2014}, which now extends to the study of spin-orbit alignment of planetary systems \citep{huberetal2013,addisonetal2018}. 

There are many examples of binaries in which both stars pulsate, which have become known as PB2s (double-pulsator binaries) in direct analogy to the SB2s of spectroscopy. PB2s can be found for many classes of pulsator, including solar-like oscillators on the main-sequence (Sect.\,\ref{sec:solar}), $\delta$\,Sct stars (Sect.\,\ref{sec:dsct}), red giants (Sect.\,\ref{sec:rg}), and $\gamma$\,Dor stars (Sect.\,\ref{sec:gdor}). Compact pulsators including hot subdwarfs and white dwarfs are yet to be found in PB2s, but we discuss their existence in binary systems in Sect.\,\ref{sec:compact}, including attempts to use pulsation timing to derive their orbits. Not all types of pulsation are suitable for the derivation of orbits by pulsation timing, since the oscillation modes must be coherent on time-scales longer than the orbital period. However, with the help of asteroseismology, mass ratios can sometimes be inferred anyway, as we will show in this review. Finally, we discuss tidally excited oscillations (TEOs) seen in short-period eccentric binaries (Sect.\,\ref{sec:TEOs}), and present a summary in Sect.\,\ref{sec:conclusions}.


\section{Main-sequence solar-like oscillators in binaries}
\label{sec:solar}

\textit{Kepler} had a short-cadence (SC) mode well-suited to studying the high oscillation frequencies of main-sequence sun-like stars \citep{gillilandetal2010a}. While dozens of apparently single pulsators were discovered \citep{chaplinetal2011a,lundetal2017}, only a handful of binaries are known in which solar-like oscillations have been observed in both components. Most of these have been studied as individual stars: 16\,Cyg\,A and B (KIC\,12069424 and KIC\,12069449; \citealt{metcalfeetal2012,metcalfeetal2015,daviesetal2015}), and
HD\,176071 (KIC\,9139151 and KIC\,9139163; \citealt{appourchauxetal2012}) are examples from \textit{Kepler}, while the $\alpha$~Cen AB system has been studied from the ground (e.g. \citealt{bouchy&carrier2002,beddingetal2004,kjeldsenetal2005,joyce&chaboyer2018}). Three binaries, however, were discovered from single light curves containing two pulsation spectra. These genuine PB2s are KIC\,7510397 \citep{appourchauxetal2015}, KIC\,10124866 \citep{whiteetal2017a} and the subgiants in KIC\,7107778 \citep{ylietal2018}.

Solar-like oscillations of main-sequence stars are not coherent enough to determine orbital elements from pulsation timing, but the oscillations still contain a lot of information. The asteroseismic scaling relations \citep{kjeldsen&bedding1995}, which can be tested using pulsating stars in binaries (see Sect.\,\ref{ssec:scaling}), can be used to estimate the mass ratio of the stars based on their frequency of maximum power, $\nu_{\rm max}$, and their large frequency spacing, $\Delta \nu$.

\begin{figure*}
	\centering
	\includegraphics[width=0.95\linewidth]{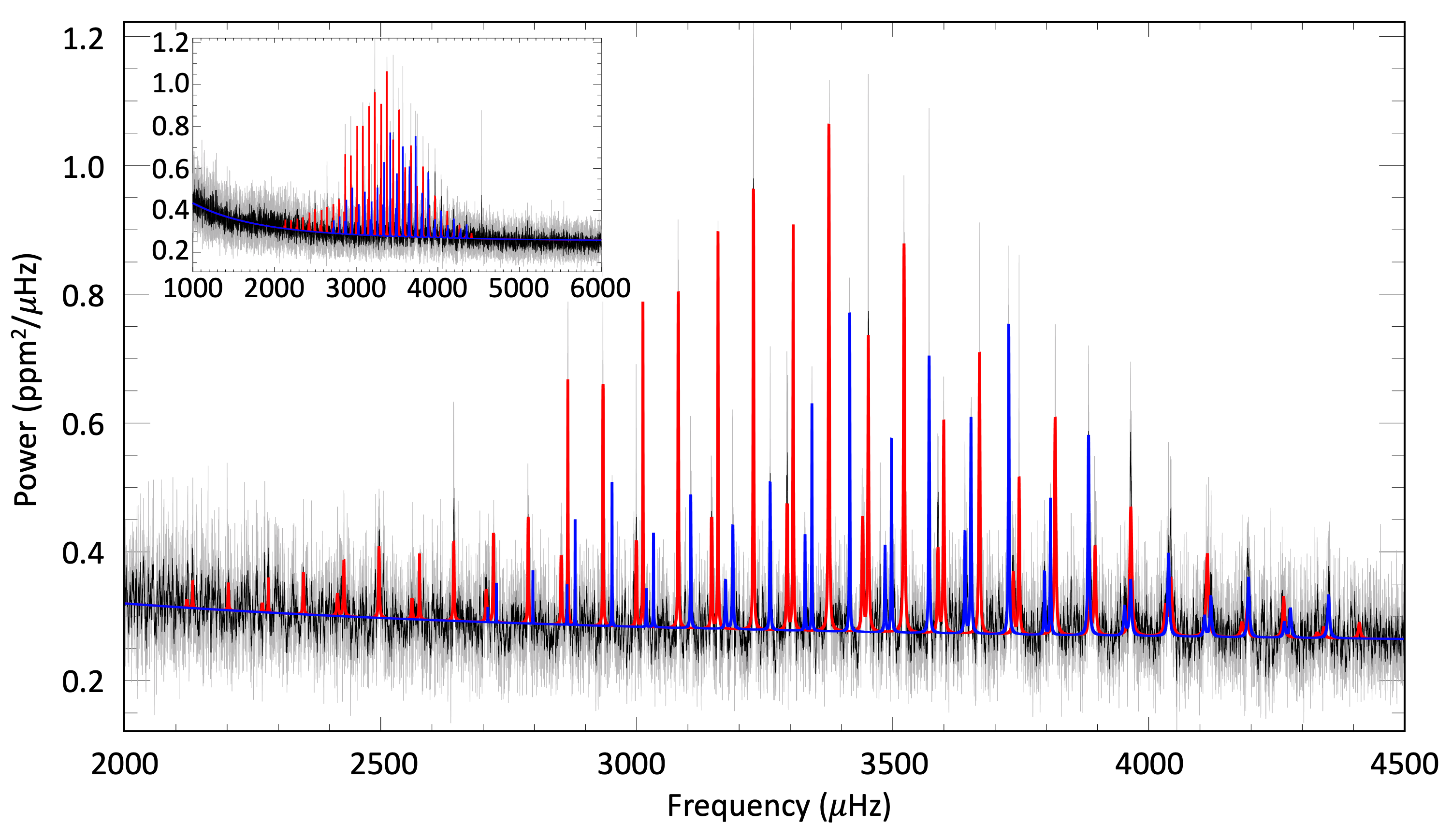}
	\caption{The PB2 KIC\,10124866, known colloquially as Luke \& Leia, consists of two solar-like oscillators with a mass ratio of $q=m_2/m_1=0.98$. The higher-mass, primary component, is shown in red. Figure modified from \citet{whiteetal2017a}.}
	\label{fig:lukeleia}
\end{figure*}

The large spacing $\Delta \nu$ scales as the square-root of the mean density of the star, $\sqrt{\bar{\rho}}$ \citep{ulrich1986,kjeldsen&bedding1995}. For a given age, higher-mass stars have lower mean densities, so more massive (more evolved) stars have a smaller $\Delta \nu$. This difference is measurable for KIC\,10124866 (Fig.\,\ref{fig:lukeleia}) -- the mode frequencies of the primary (red) are closer together than those of the secondary (blue),  such that the primary has one more radial order of pulsation between 2700 and 4000\,$\mu$Hz. A further consequence of the mass difference is that $\nu_{\rm max}$ is 7\% lower for the more massive component, and its pulsation amplitude is higher. The latter observation has two causes: (i) the mass-luminosity relation is steep, so the 2\% difference in mass leads to a $\sim$7\% difference in luminosity, hence the oscillations of the less-massive star are more diluted in the light curve; and (ii) more evolved stars have higher pulsation amplitudes \citep{yuetal2018}. Further details on this binary can be found in \citet{whiteetal2017a}.

\citet{ylietal2018} studied the pair of subgiants in KIC\,7107778. Their masses differ by only 1\%, so the oscillation spectra still overlap even in the subgiant phase. Binaries like this are rare. The fraction of twins (mass ratio $q>0.95$) at the masses and orbital period of KIC\,7107778 ($1.4$\,M$_{\odot}$, 39\,yr; \citealt{ylietal2018}) is no more than 10\% \citep{moe&distefano2017}. Finding a pair of stars with a mass ratio so close to unity, and in a rapid evolutionary phase, is fortunate indeed.

\begin{figure}
	\centering
	\includegraphics[width=0.95\linewidth]{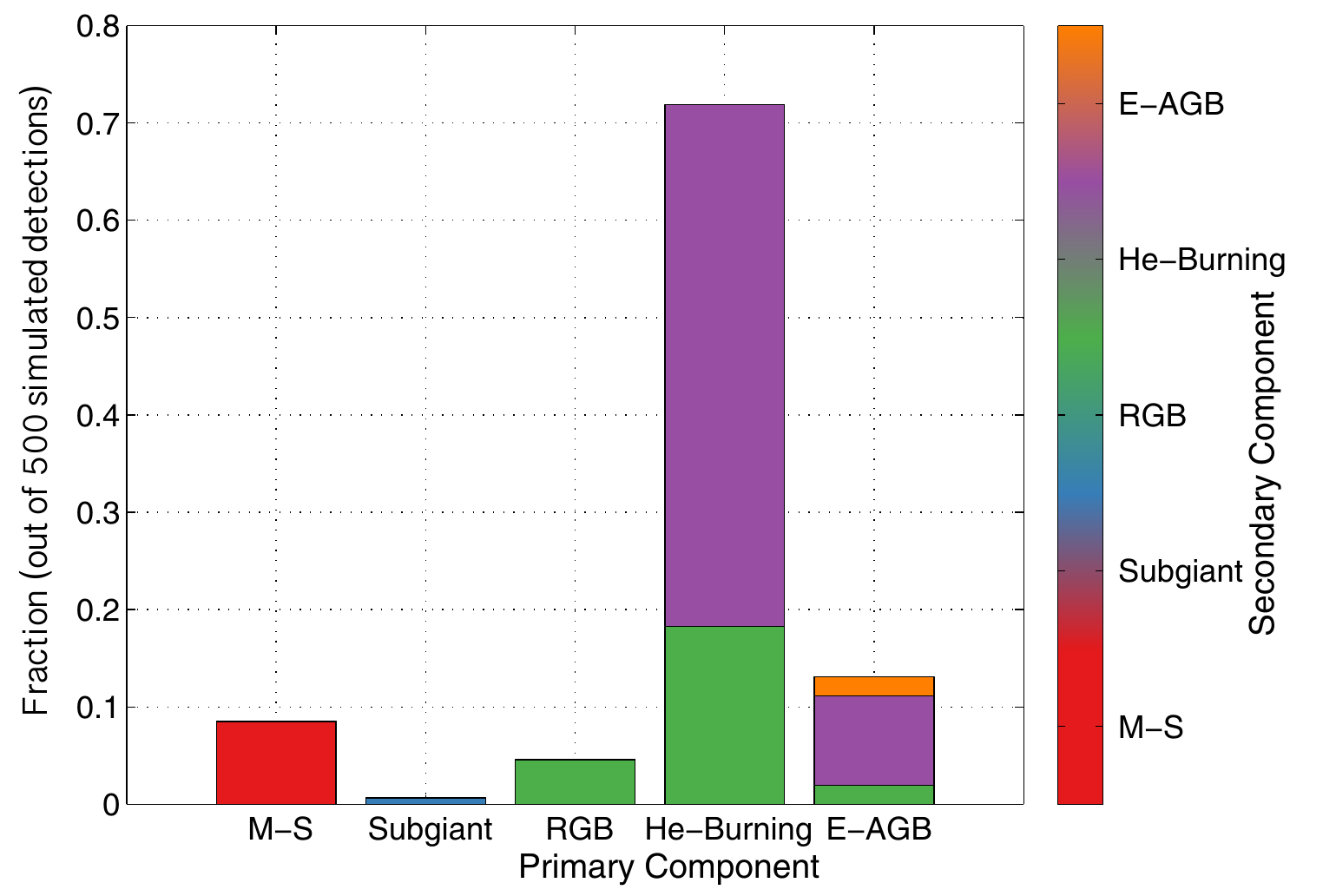}
	\caption{Detectable asteroseismic binary pairs from a simulation of the \textit{Kepler} field. In most of the 500 detectable binaries, one or both components are He-burning giants. Figure modified from \citet{miglioetal2014}.}
	\label{fig:binary_sim}
\end{figure}

\citet{miglioetal2014} performed simulations on the \textit{Kepler} field to determine how many solar-like oscillators in PB2s should be evident in \textit{Kepler} data. They further categorised them according to the evolutionary stage of the primary and secondary components (Fig.\,\ref{fig:binary_sim}). The detectability of PB2s depends on many factors, such as the population of Milky Way stars probed by the \textit{Kepler} field, especially their masses and ages \citep{girardi2016,sharmaetal2016,sharmaetal2017}; how binary properties such as the mass ratio (and hence luminosity ratio) vary according to the underlying population statistics \citep[][and references therein]{moe&distefano2017}; and how these properties evolve as the stars age. The detectability of binaries also hinges on the detectability of the pulsations in the \textit{Kepler} data \citep{ballotetal2011,chaplinetal2011}. The lower amplitudes of main-sequence stars and dilution by the luminosity ratio are additional variables. Only $\sim$40 detectable main-sequence pairs are therefore expected in the \textit{Kepler} field \citep{miglioetal2014}, and this number is further reduced because of the limited availability of \textit{Kepler} SC slots -- only 512 targets could be observed in SC mode at any given time \citep{gillilandetal2010b}. Only 1 or 2 solar-like PB2s with a subgiant component are expected in the \textit{Kepler} field; those found by \citet{ylietal2018} are probably the only ones. We will return to look at the more common red giant binaries in Sect.\,\ref{sec:rg}.

\section{$\delta$\,Sct stars}
\label{sec:dsct}

We pause our discussion of solar-like oscillators to appreciate the power of pulsation timing in providing binary orbits. No class of pulsators has seen more successful application of this method in \textit{Kepler} data than $\delta$\,Sct stars. The number of pulsation-timing binaries exceeds even the number of eclipsing binaries in \textit{Kepler} data of stars in and around the $\delta$\,Sct instability strip \citep{kirketal2016,murphyetal2018}.

\subsection{Orbits and pulsation timing}

The pulsation modes of $\delta$\,Sct stars are very stable in frequency, with few exceptions. \citet{bowmanetal2016} found no evidence in \textit{Kepler} data for intrinsic phase modulation (equivalent to frequency modulation) in 986 stars examined, unless that phase modulation was accompanied by amplitude modulation due to beating of pulsation modes. Ground-based monitoring of 4\,CVn concluded that mode frequencies are stable on timescales of at least a few decades \citep{breger2000b}. Main-sequence evolution causes period increases of only ($1/P$ d$P$/d$t$ $=$) $10^{-10}$\,yr$^{-1}$, hence the pulsation modes are excellent stellar clocks.

With 4\,yr of continuous data, the pulsation frequencies of \textit{Kepler} $\delta$\,Sct stars can be measured very precisely, and their phases can be monitored for periodic shifts due to binary motion. As a pulsating star moves in its orbit, the path length between the star and the telescope changes,\footnote{The motion of the telescope around the solar system barycentre must be corrected for as well (\citealt{shibahashi&murphy2018}, these proceedings).} which causes the pulsation maxima and minima to arrive early or late. These arrival-time delays (early arrival is simply a negative delay) give the light travel time across the orbit \citep{murphyetal2014,murphy&shibahashi2015}, i.e. its physical size.

\begin{figure}
	\centering
	\includegraphics[width=0.95\linewidth]{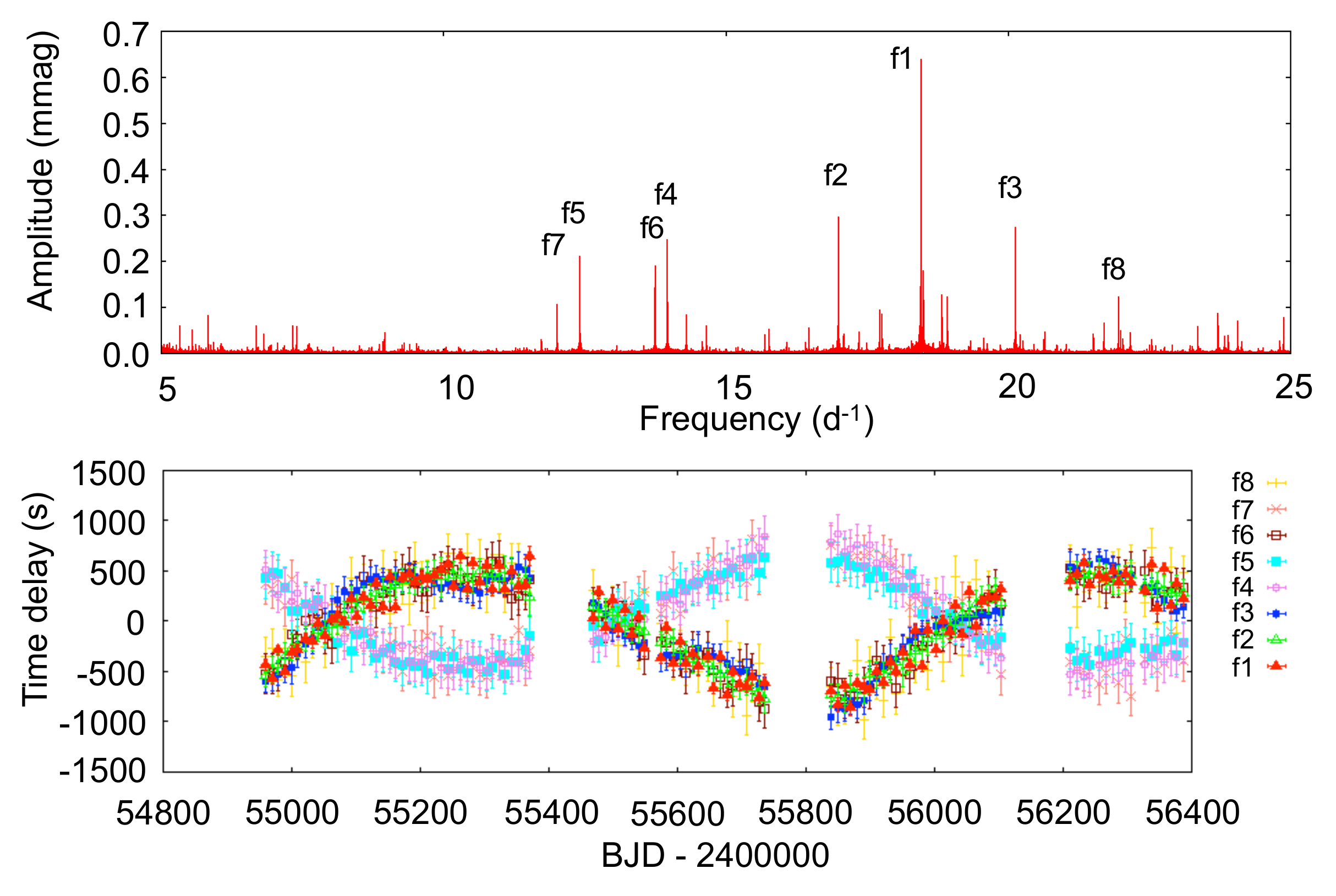}
	\caption{The $\delta$\,Sct PB2 system, KIC4471379. The top panel shows the Fourier transform of the \textit{Kepler} light curve with some of the strongest mode frequencies labelled, and the lower panel shows time delays for those modes. The time delays of the two stars are clearly in anti-phase, as expected for two stars on opposing sides of the barycentre. Figure modified from \citet{murphyetal2014}.}
	\label{fig:dsct_pb2}
\end{figure}

Over 90\% of $\delta$\,Sct binaries detected with this pulsation-timing method are PB1s (single-pulsator binaries), where no pulsation is evident from the companion \citep{murphyetal2018}. This is mainly because the mass-ratio distribution of binaries with A-type primaries peaks at $q\sim0.2$ \citep{murphyetal2018}, so most companions to these 1.5--2.5\,M$_{\odot}$ $\delta$\,Sct stars are 0.3--0.5\,M$_{\odot}$ K/M dwarfs. Around 20\% of the detected companions are actually white dwarfs, but no pulsation attributed to the white-dwarf component has yet been reported to make these PB2s.
Another reason for the low fraction of PB2s is that a detection bias exists against them, because the pulsation timing signal from each component partially cancels out. Nonetheless, 24 PB2s of $\delta$\,Sct pairs are known from \textit{Kepler} data (an example is shown in Fig.\,\ref{fig:dsct_pb2}), and $\theta^2$\,Tau is a good case study from the ground that also has a spectroscopic and astrometric orbit \citep{bregeretal2002b,armstrongetal2006,torresetal2011}.

The large number of pulsation-timing binaries allows for statistical analyses to be conducted. In addition to mass-ratio distributions, it is possible to investigate both the eccentricity distribution and binary fraction of these A/F primaries \citep{murphyetal2018}, which give clues about binary star formation mechanisms \citep{tohline2002,kratter&matzner2006,kratteretal2010a}. The binary fractions can be compared to other regimes of primary mass \citep{moe&distefano2017,murphyetal2018} to understand multiplicity as a function of mass, or compared to surveys in RV \citep[e.g.][]{lampensetal2017}, adaptive optics \citep[e.g.][]{derosaetal2014}, long-baseline interferometry \citep[e.g.][]{rizzutoetal2013} and other techniques to understand the binary fraction as a function of orbital period.

Asteroseismology is a major beneficiary of binary analyses because there remains much to learn about mode selection and excitation in $\delta$\,Sct stars. They are generally rapid rotators with mean equatorial velocities in excess of 150\,km\,s$^{-1}$ \citep{royeretal2007}, which makes precise spectroscopic inference difficult. Spectral lines are broad, so metallicities cannot be determined precisely. Even within clusters, comparisons between unrelated stars are complicated by large star-to-star abundance anomalies \citep{gebranetal2010}. The p\,modes in $\delta$\,Sct stars do not yet lend themselves to determining ages precisely. All of these obstacles might be overcome by comparative asteroseismology between pulsating stars in binary systems, which could reduce the number of free parameters enough to get a foothold on mode identification, but breakthroughs have not been forthcoming.

\subsection{Eclipsing systems containing $\delta$\,Sct stars}

The greatest asteroseismic potential lies in the eclipsing binaries (EBs), where the orbital inclination and the ratio of the stellar radii to the semi-major axis (i.e. $R_1/a_1$ and $R_2/a_2$) can be measured. Ideally, the eclipse data are supplemented with dynamical masses from another method, such as RV monitoring. This allows the absolute masses (with no $\sin i$ dependency) to be derived because $i$ can be measured. With a period and masses the semi-major axis becomes known, subsequently allowing absolute radii (not convolved with the semi-major axis) to be measured. Pulsation timing can substitute for or complement RVs to provide these dynamical masses for the eclipsing systems, but this requires accurate eclipse modelling, so the EBs were set aside from the largest pulsation timing catalogue \citep{murphyetal2018}.

Many $\delta$\,Sct stars in EBs are known, approximately 40 of which were discovered in \textit{Kepler} data \citep{guzik&gaulme2014,kirketal2016,murphyetal2018}. However the largest collection comes from ground-based studies, as catalogued recently by \citet{liakos&niarchos2017}. They compiled over 100 systems, with only small overlap with the \textit{Kepler} targets. It appears from their sample that binarity weakly influences the pulsation periods, for orbital periods below $\sim$13\,d, though it is not clear whether this is directly a result of tidal interaction or if mass transfer has played a role. The large scatter around the $P_{\rm orb}$--$P_{\rm puls}$ `relation' casts doubt on its usefulness; these targets should be revisited with {\it TESS} to see if more can be learned.

Knowledge of the stellar inclination greatly empowers asteroseismology. The visibility of non-radial modes depends on the degree $\ell$, the azimuthal order $m$, and the stellar inclination \citep{dziembowski1977,gizon&solanki2003,aertsetal2010}. A well-constrained inclination can therefore aid mode identification by eliminating the low-visibility modes from the many possibilities, adding further appeal to the study of pulsators in EBs.

Inclinations from EBs can also lessen the difficulties in mode identification caused by stellar rotation. In a non-rotating star, modes of different $m$ all have the same frequency if $\ell$ and the radial order $n$ are the same, but rotation lifts this degeneracy \citep{ledoux1951}. Slow rotators have pulsation modes split by the rotation frequency, which actually aids mode identification, but in moderate rotators the second order effects become important \citep{saio1981}. Rapid rotators, on the other hand, are so aspherical that pulsation modes become confined to equatorial regions and adopt complicated characteristics \citep{lignieres&georgeot2009,reeseetal2013}. Thus, when the star rotates at least `moderately', with $v\sin i \gtrsim 50$\,km\,s$^{-1}$, as the majority of A and F stars do \citep{royeretal2007}, the pulsation spectrum becomes complicated enough that mode identification becomes intractable. Knowing the inclination alleviates this, because extracting the equatorial rotation rate from $v\sin i$ allows the spherical distortion to be quantified, and frequency splittings between modes of the same $n$ and $\ell$ can be predicted.

The relationship between eclipses and rotation goes deeper. EBs are generally close binaries, because eclipses become geometrically unlikely when the semi-major axis is large. In close binaries, tidal effects are strong enough to rotationally brake the stars \citep{zahn1975,zahn1977,tassoul&tassoul1992a}, especially if tidally excited oscillations are present \citep{witte&savonije2001,fuller2017}. This aids the study of $\delta$\,Sct stars in two ways: (i) the rotation and orbital periods tend towards synchronicity, so if the former is unable to be measured by other means then it can often be assumed to equal the latter (however, the effectiveness of tidal synchronisation depends on the stellar masses and evolutionary states; see the caveats in Sections\,\ref{sec:rg} and \ref{sec:compact}); and (ii) the rotation rate is slower, which simplifies the pulsation spectrum. The full interaction between tides and pulsation remains an active area of study.

Finally, eclipses function as a spatial filter, temporarily masking part of the pulsating star from view. This filter can be used to identify pulsation modes if the pulsation amplitudes can be monitored as a function of stellar longitude and latitude. This method has enjoyed only limited success with $\delta$\,Sct stars so far \citep{gamarovaetal2003,mkrtichianetal2004}.

\subsection{Selected case studies of $\delta$\,Sct stars in eclipsing systems}

\subsubsection{Time delays of a singly eclipsing system}

\begin{figure}
	\centering
	\includegraphics[width=0.95\linewidth]{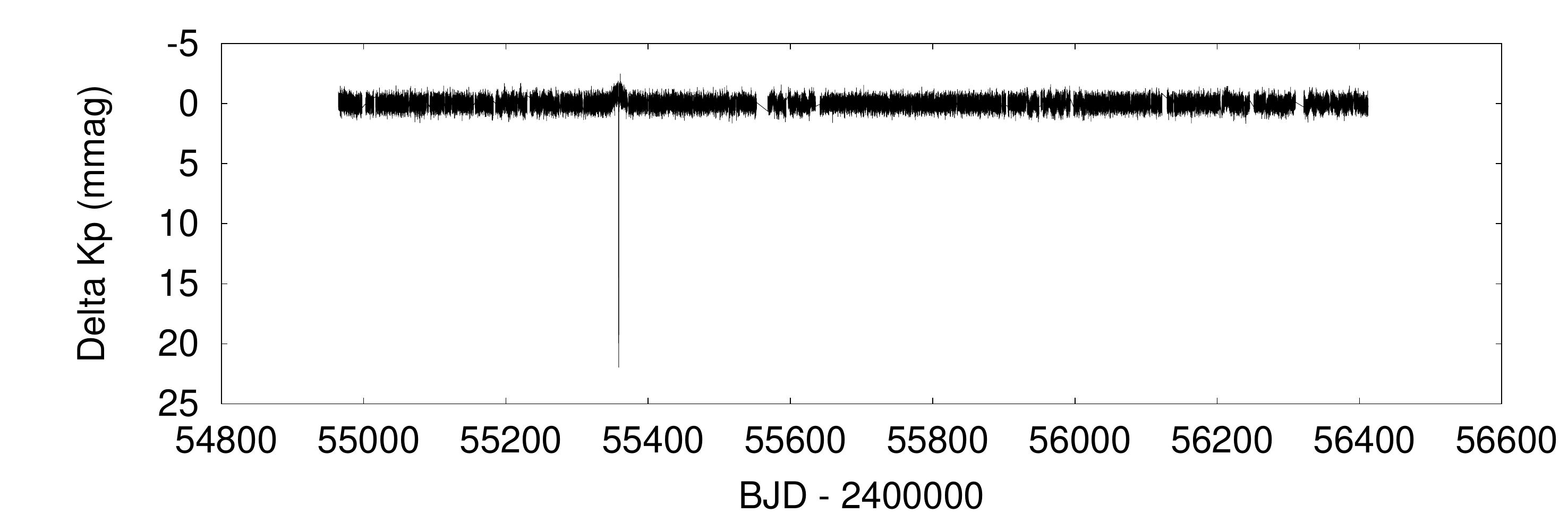}
	\includegraphics[width=0.95\linewidth]{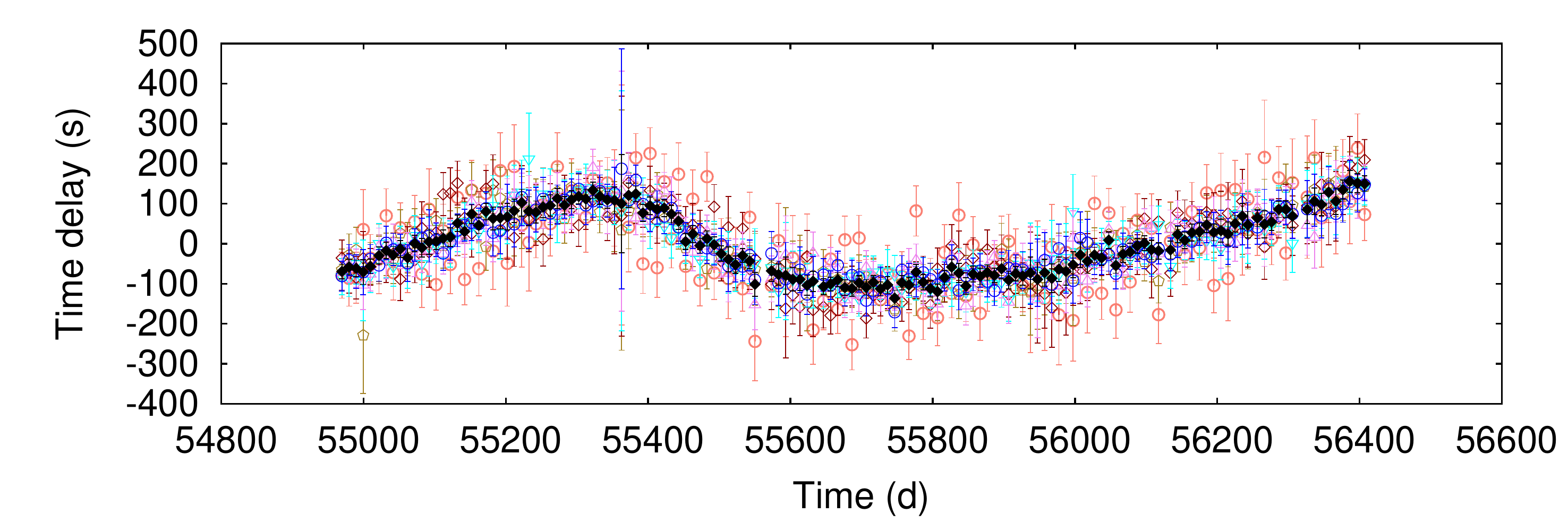}
	\caption{The light curve (top) and time delays (bottom) of the singly-eclipsing system KIC\,8648356.}
	\label{fig:single_eclipse}
\end{figure}

KIC\,8648356 has just a single eclipse in the entire 4-yr \textit{Kepler} data set (Fig.\,\ref{fig:single_eclipse}). The eclipse occurs at time-delay maximum, which tells us that the $\delta$\,Sct star is being eclipsed, however no corresponding eclipse occurs at time-delay minimum, even in the residuals after prewhitening pulsation frequencies. Identifying which star is pulsating in this manner is trivial with a time-delay orbit. The time-delay curve suggests that another primary eclipse was due just after \textit{Kepler} ceased observation of its original field.

\subsubsection{The multi-eclipsing system KIC\,4150611}

\begin{figure}
	\centering
                \begin{overpic}[width=0.95\linewidth]{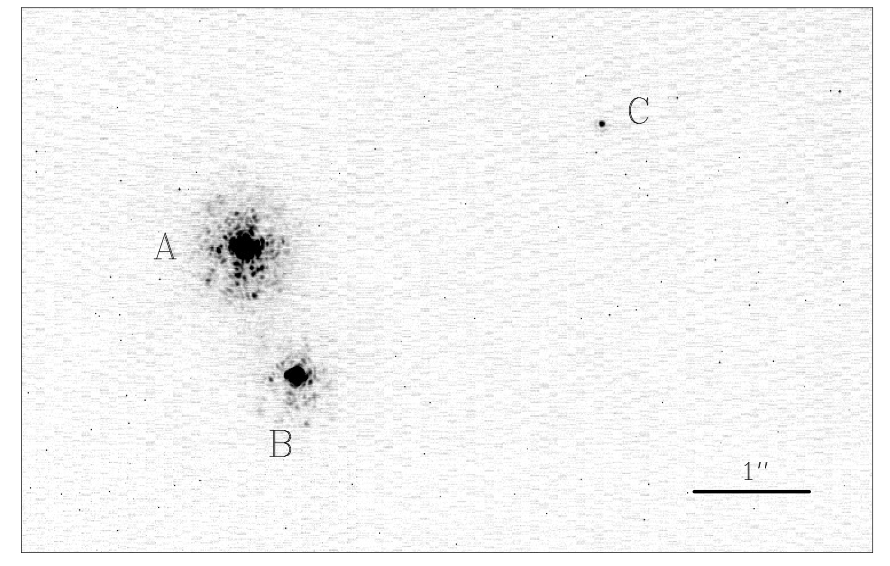}
                \put(66,44){\blu{\small ? + ?}}
                \put(37,15){\blu{\small G + G}}
                \put(15,50){\blu{ \small $\delta$\,Sct / $\gamma$\,Dor hybrid}}
                \put(15,45){\blu{ \small triple: F1 + M + M}}
                \end{overpic}
	\includegraphics[width=0.95\linewidth]{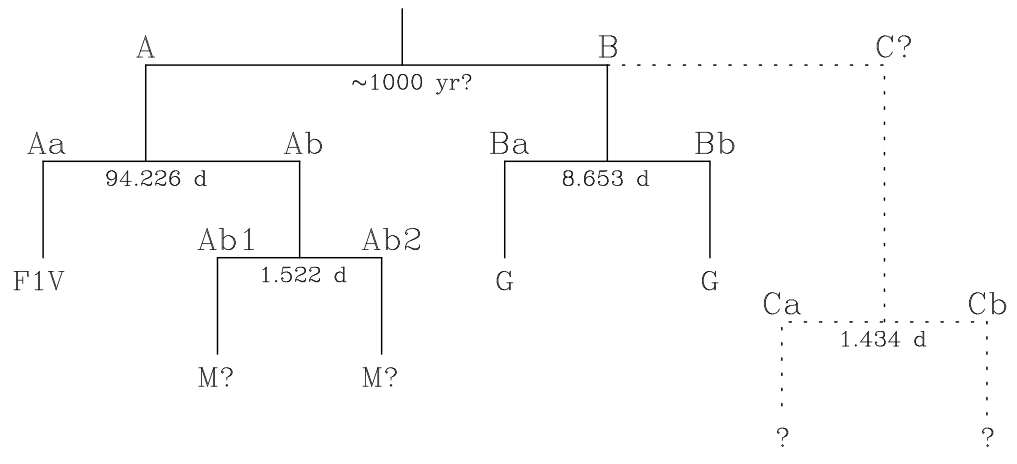}
	\caption{Top: NIRC2 (Keck) image of KIC\,4150611. North is up, east is left. Modified from \citet{helminiaketal2017}. Bottom: ``mobile diagram'' of the KIC\,4150611 system, from \citet{helminiaketal2017}.}
	\label{fig:4150611_system}
\end{figure}

KIC\,4150611 is a bright (V=8) quintuple system with a $\delta$\,Sct/$\gamma$\,Dor hybrid as the primary. As companions, it has a pair of M dwarfs that eclipse each other every 1.52\,d, and a pair of G dwarfs that eclipse each other every 8.65\,d \citep{helminiaketal2017}. Four periods of eclipse are observed in the \textit{Kepler} light curve; the third occurs when the M dwarfs eclipse the primary every 94.23\,d, and the fourth, of 1.43-d period, is tentatively assigned to a distant component that might be gravitationally bound. The relationship between them in shown in Fig.\,\ref{fig:4150611_system}. Among the remarkable features of this system is that the pair of M dwarfs eclipse each other whilst eclipsing the F1 primary.

\section{Red giants}
\label{sec:rg}

\subsection{Pulsation Timing}
\label{ssec:rg_timing}

\citet{comptonetal2016} investigated the detectability of binaries by pulsation timing for a variety of pulsation classes and orbit sizes. It was hoped that the mixed modes of red giants, which have long lifetimes compared to their pure p-mode counterparts, might offer similar promise to $\delta$\,Sct stars. Unfortunately the oscillation periods of red giants are too long and their frequency spectra too crowded to have enough sensitivity and resolution to find binary companions.

Red giants have plenty to offer binary analyses in other ways. Dynamical masses are usually convolved with $\sin i$ because inclination angles, $i$, are generally unknown. Hence only minimum masses are measured, and even in PB2s only the mass ratio and minimum masses are known. However, masses for red giants are available from the asteroseismic scaling relations, breaking this degeneracy and allowing the masses of both stars to be determined. An excellent application of this was in the analysis of 18 pulsating red giants in eccentric binaries by \citet{becketal2014}. In one of those it was possible to infer the rotation period of the red giant, which surprisingly was not pseudo-synchronous with the orbital period.

\begin{figure}
	\centering
\begin{overpic}[width=0.95\linewidth]{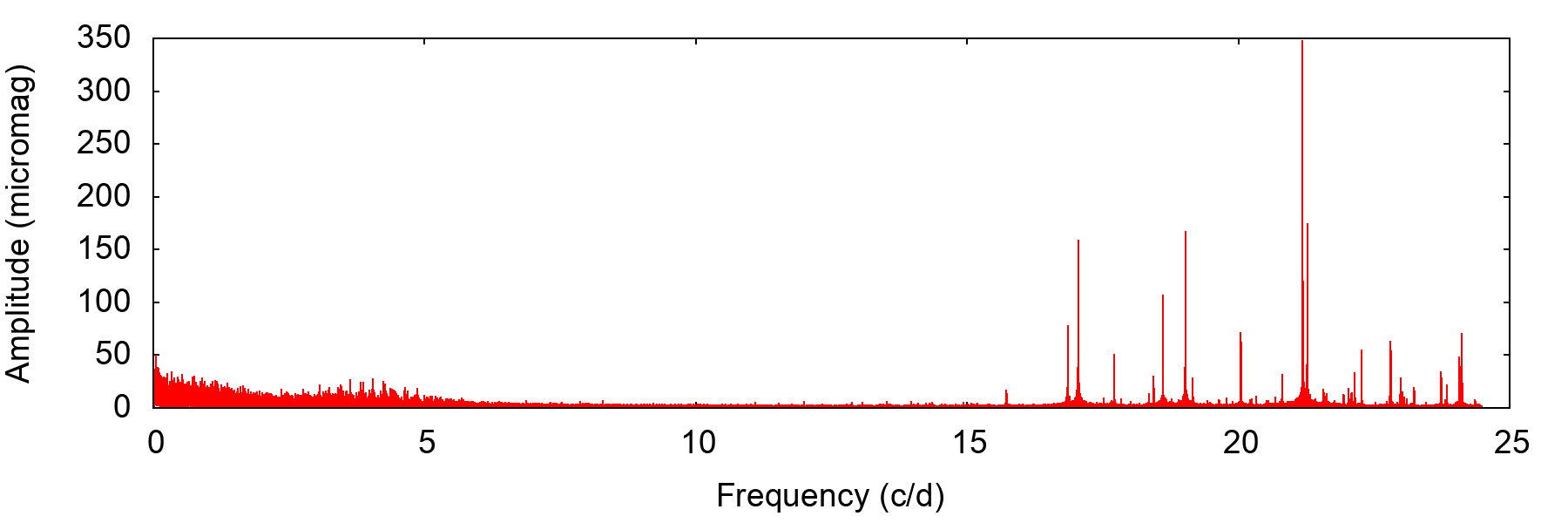}
\put(90,27){\small (a)}
\put(15,27){\small KIC\,6753216}
\end{overpic}
\begin{overpic}[width=0.95\linewidth]{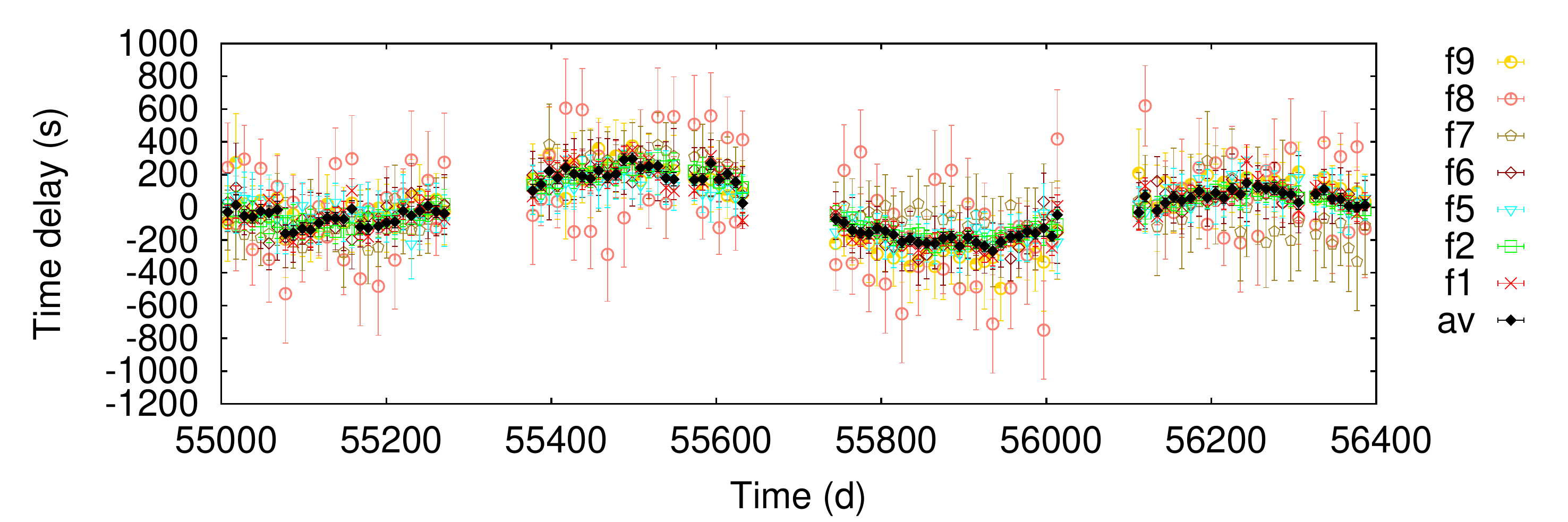}
\put(82,27){\small (b)}
\end{overpic}
	\caption{A red-giant + $\delta$\,Sct pair in \textit{Kepler} data. The Fourier transform of the \textit{Kepler} light curve {\bf (a)} shows red-giant oscillations at 3--5\,d$^{-1}$ (33--55\,$\upmu$Hz), along with a granulation signature at lower frequency. The narrow peaks above 15\,d$^{-1}$ are $\delta$\,Sct p\,modes, some of which lie above the \textit{Kepler} long-cadence Nyquist frequency (24.4\,d$^{-1}$). Panel {\bf (b)} shows the time delays of the $\delta$\,Sct pulsations, showing that this is a binary system.}
	\label{fig:rg_dsct}
\end{figure}

Pulsation timing is still possible for red giant orbits if the companion has coherent oscillations. Good candidates for this would be $\beta$\,Cep or $\delta$\,Sct stars, because they are luminous enough to be seen against the bright red giant. But \textit{Kepler} observed few B stars, and $\delta$\,Sct companions to red giants are very rare. Just one, KIC\,6753216, was recorded by \citet{colmanetal2017} in their search for close companions to red giants. Their pixel analysis suggests the stars in KIC\,6753216 are physically associated, and the $\delta$\,Sct pulsations show a clear time-delay signal (Fig.\,\ref{fig:rg_dsct}) that indicates this is a 760-d binary (Murphy, Colman et al. in prep.).

\subsection{Testing Asteroseismic Scaling Relations}
\label{ssec:scaling}

The asteroseismic scaling relations are a crucial toolkit for obtaining stellar parameters ($R$, $M$) from pulsation properties ($\nu_{\rm max}$, $\Delta \nu$). They can be applied quickly and simply to large samples of stars, and have proved indispensable to the analysis of the 16\,000 oscillating red giants observed by \textit{Kepler} \citep[][and references therein]{yuetal2018}. Thousands more red giants have also been observed by {\it K2}, paving the way for archaeology of the Milky Way \citep{stelloetal2017}.

The scaling relations operate on the assumption that stars are homologous, which seems to hold well on the main sequence. However, structural changes that develop between the TAMS and the red clump might break that assumption, thence calling asteroseismically derived masses and radii into question. Eclipsing binaries offer model-independent measurements against which the scaling relations can be tested.

In close binaries, where the semi-major axis, $a$, is small enough for tidal locking, red-giant oscillations are suppressed \citep{gaulmeetal2016,rawls2016}. However, wide orbits with $a> 5 R_{\rm RG}$ make good benchmarks against which the scaling relations appear imperfect \citep{rawls2016,kallingeretal2018,themessletal2018}.

Corrections to the scaling relations have been proposed. \citet{kallingeretal2018} proposed the introduction of empirically-derived non-linear terms, whereas \citet{themessletal2018} found the linear scaling relations to be sufficient when adopting a different empirical reference $\Delta \nu$ for the Sun (see also \citealt{guggenbergeretal2016}). \citet{vianietal2017} suggested the inclusion of a molecular weight term to the $\nu_{\rm max}$ equation. These developments are too new for a unified correction procedure to have been agreed upon.

The scaling relations appear to be very accurate when empirical or theoretical corrections are applied \citep{handbergetal2017,brogaardetal2018}. Including other asteroseismic parameters, such as the period spacing of gravity modes $\Delta P$, can further improve the accuracy of asteroseismic results \citep{rodriguesetal2017}. It remains the case that, when the highest accuracy is required, there is no substitute for detailed frequency modelling. It is by detailed frequency modelling that \citet{tlietal2018} were able to calibrate the mixing length and asteroseismic surface term using red giants in eclipsing binaries.

\section{$\gamma$\,Dor stars}
\label{sec:gdor}

The study of $\gamma$\,Dor stars has seen remarkable progress in the space era. Their pulsation spectra are often dense, with dozens of modes near 1\,d$^{-1}$ and period spacings of $10^3$\,s ($\sim$0.01\,d). The mode density is higher in slowly rotating $\gamma$\,Dor stars, where small rotational splittings are observed \citep[e.g.][]{kurtzetal2014,saioetal2015}. To measure these requires excellent frequency resolution, demanding datasets of $\gtrsim$1\,yr duration. Additionally, $\gamma$\,Dor stars are difficult to study from the ground because their frequencies range between ~0.3 and 3\,d$^{-1}$ \citep{vanreethetal2015b,glietal2018}. For both reasons, continuous observations from space with long time spans have revolutionised our understanding.

All $\gamma$\,Dor stars pulsate in dipole ($\ell=1$) gravity modes, and the slowly rotating ones can pulsate in all three azimuthal orders ($m=0,\pm1$). Rotation modifies the period spacings between consecutive radial orders, and imparts a gradient (`slope') on the period spacings that depends on the azimuthal order \citep{ouazzanietal2017}. More rapid rotators have steeper slopes, with smaller dependence on the excited radial orders, hence the slopes are a useful diagnostic of the rotation when rotational splittings are too large to be measured. The rapid rotators can also pulsate in Rossby modes (r~modes), which help constrain the rotation rate (\citealt{vanreethetal2016,saioetal2018a}; Li et al. in prep), chemical gradients in the interior, and any radial differential rotation \citep{vanreethetal2018}. Importantly, both the slope and the frequency splittings give the rotation {\it rate} (not velocity) of the near-core region (not the surface).  Since the surface is readily probed by other means, namely spectroscopic $v\sin i$ or p-mode frequency splittings in $\gamma$\,Dor--$\delta$\,Sct hybrids, then measurements of the near-core rotation rate shed light on the internal rotation profile of the star \citep[e.g.][]{schmid&aerts2016,murphyetal2016a,vanreethetal2018}, and thence angular momentum transport \citep{aerts2015,aertsetal2018b}.

\begin{figure}
	\centering
\begin{overpic}[width=0.99\linewidth]{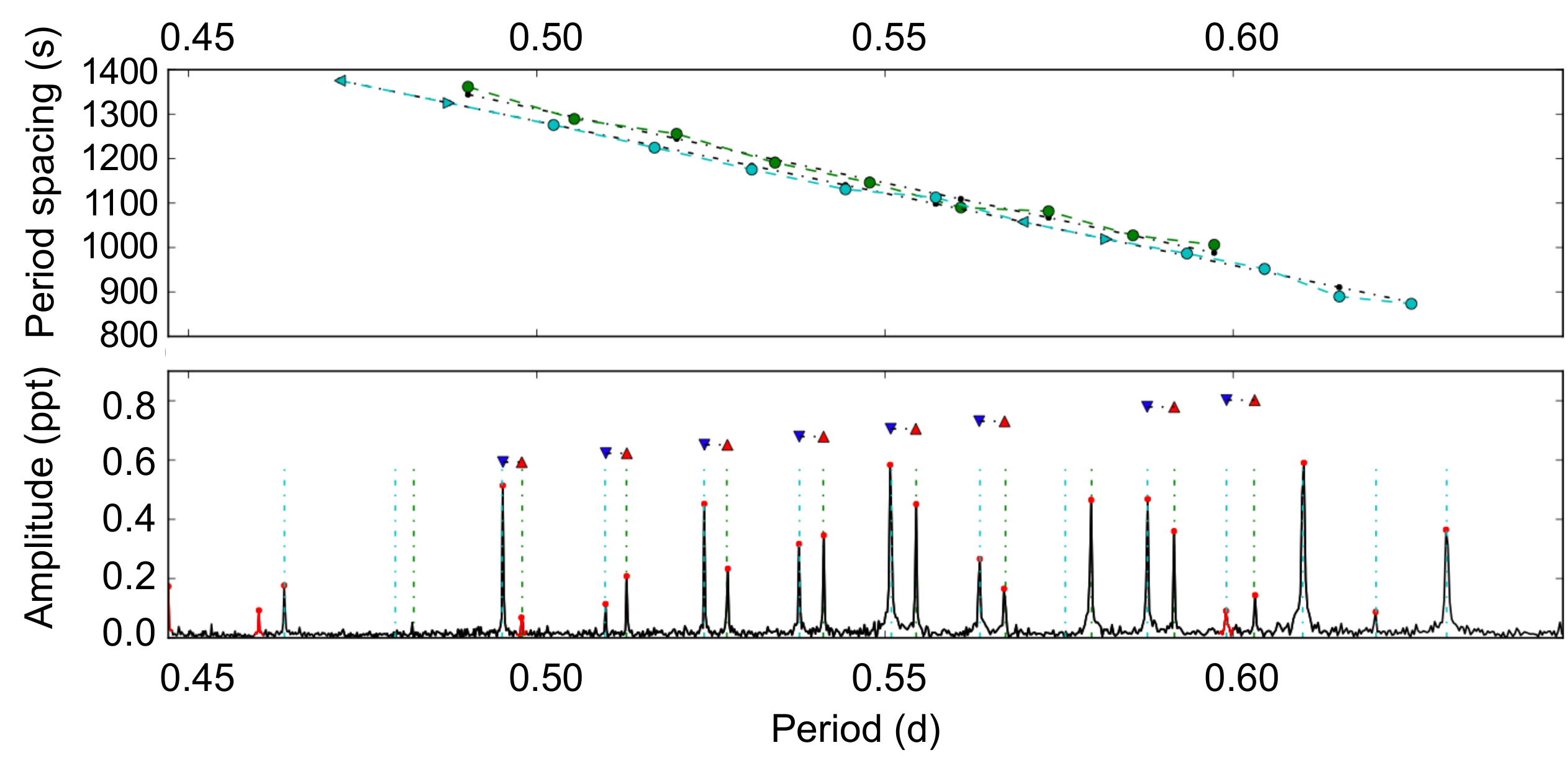}
\put(94,41){\small (a)}
\put(94,22){\small (b)}
\end{overpic}
	\caption{Period spacings {\bf (a)} for two series of dipole modes in the periodogram {\bf (b)} of the $\gamma$\,Dor PB2 KIC\,6862920. The period spacings overlap and have similar gradients, hence cannot be from a single star. Modified from \citet{glietal2018}.}
	\label{fig:gdorPB2}
\end{figure}

The dense pulsation spectra of $\gamma$\,Dor stars make them poorly suited to pulsation timing, but they are detectable as PB2s. \citet{glietal2018} found two PB2s based on their slopes and frequency splittings. The splittings were small, suggesting long rotation periods, but the slopes were steep, suggesting short rotation periods. The period spacing patterns were peculiar for having similar gradients and for almost overlapping (Fig.\,\ref{fig:gdorPB2}), which for single stars contradicts current theory. In each light curve, the two period spacing patterns were originating from two different $\gamma$\,Dor stars. They also happened to be $\delta$\,Sct stars (i.e. hybrids), where pulsation timing on the p~modes gave mass ratios and orbital parameters \citep{murphyetal2018}.

The best-studied example of a PB2 $\gamma$\,Dor system is the pair of $\delta$\,Sct--$\gamma$\,Dor hybrids in KIC\,10080943 \citep{keenetal2015,schmidetal2015,schmid&aerts2016,johnstonetal2019}. The short orbital period (15\,d) made pulsation timing of the p~modes difficult, but made RV analysis easy, and the orbit was seen in the light curve itself when folded on the orbital period (\citealt{schmidetal2015}; see also \citealt{shporer2017}). \citet{schmid&aerts2016} were able to model the two stars, using an equal-age and equal-composition requirement as well as the binary and pulsation properties as constraints. From this, they were able to determine diffusive mixing and convective overshoot parameters for the near-core region -- parameters that are otherwise only estimable to order of magnitude precision. This system in particular highlights the power of asteroseismology on binary stars.

\section{Compact pulsators}
\label{sec:compact}

\begin{figure}
\centering
\includegraphics[width=0.99\linewidth]{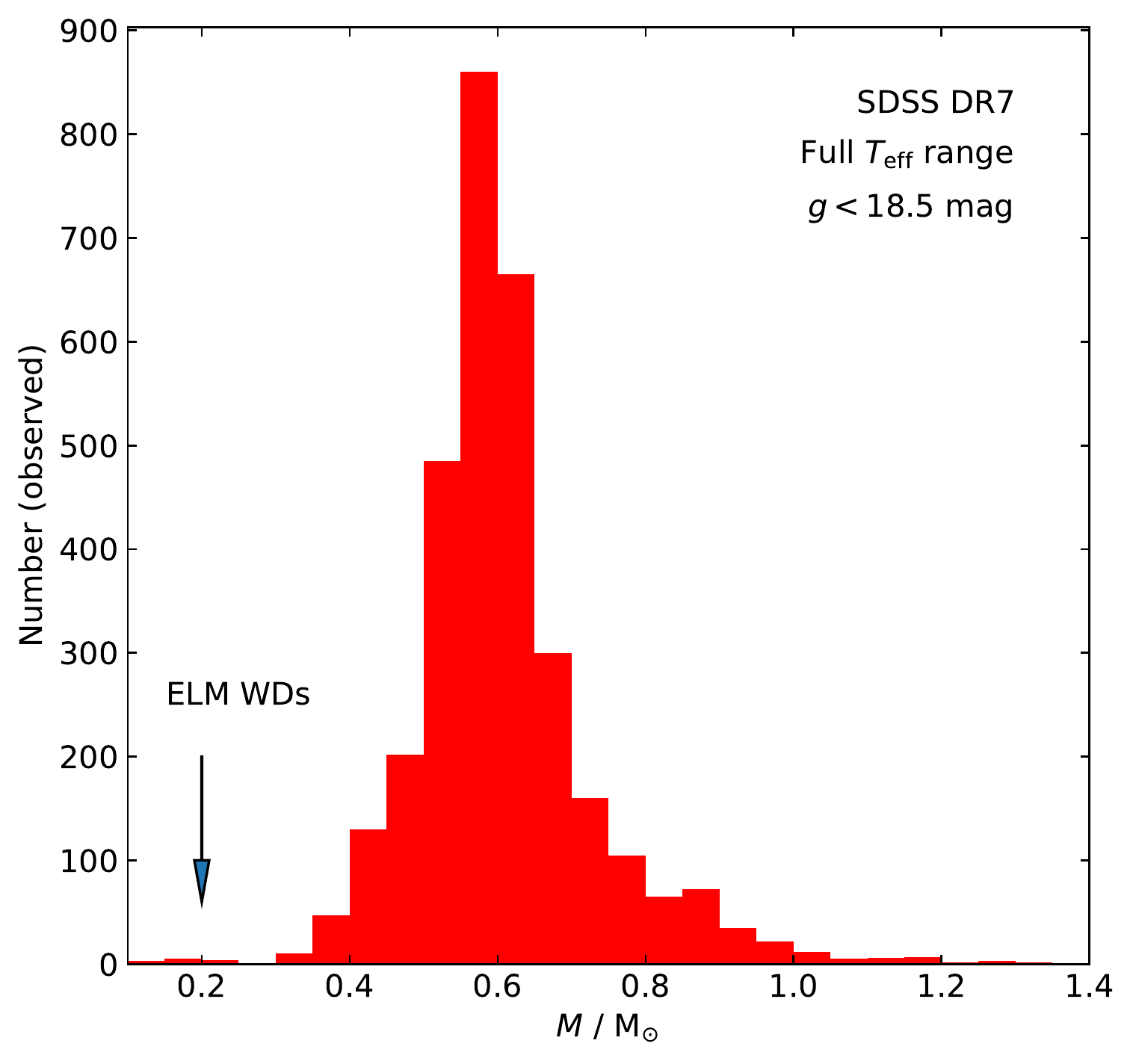}
\caption{The mass distribution of non-magnetic WDs from SDSS DR7. Extremely Low Mass (ELM) WDs are a small population with masses near 0.2\,M$_{\odot}$. Modified from \citet{tremblayetal2016}.}
\label{fig:wd}
\end{figure}

Masses for some 3000 compact remnants in SDSS DR7 have been precisely determined \citep{tremblayetal2016}. Their distribution is shown in Fig.\,\ref{fig:wd}, revealing that most are CO white dwarfs (WDs) with masses between 0.5 and 1.1\,M$_{\odot}$. Below 0.45\,M$_{\odot}$, the WDs are He-core objects that presumably originate from binary evolution, since the single-star progenitors of such low mass WDs have main-sequence lifetimes longer than the present age of the universe \citep{marshetal1995,zhaoetal2012,cummingsetal2018}. Among them are two classes of pulsators: the extremely low mass (ELM) WDs \citep{hermesetal2013d}, and the hot subdwarfs which are not shown in Fig.\,\ref{fig:wd} \citep{heber2009}. Asteroseismology offers exciting prospects for discerning their interior structure, composition, and cooling ages.

\subsection{Extremely Low Mass White Dwarfs}

ELM WDs are believed to be the originally less massive star of a close binary system \citep{sun&arras2018}. While the original primary ascends the red-giant branch, mass transfer begins and the system enters a common-envelope phase. The envelope is eventually ejected, leaving a He-/CO-core WD whose companion is still on the main sequence. When this companion ascends the red-giant branch later, a second mass-transfer episode begins, after which it is left with insufficient mass to fuse He to C and O and it becomes an ELM WD.

This precise model from \citet{sun&arras2018} predicts a narrow range of ELM WD masses ($0.146 \leq M$/ M$_{\odot}$ $\leq$ 0.18) with orbital periods between 2 and 20\,hr, matching the empirical masses, mass-ratios and orbital period distributions of ELM WD binaries \citep{boffin2015,brownetal2016}. The mass determinations are made possible by a combination of asteroseismology, spectroscopy, and binary analysis \citep{hermesetal2013d,hermesetal2013a,kilicetal2015}. One ELM WD orbits a millisecond pulsar, PSR J1738+0333, which has unusually narrow pulses, and whose pulse timings record a Shapiro delay \citep{shapiro1964}. Together, these have allowed the masses of both components to be measured to exquisite precision \citep{jacobyetal2003,jacobyetal2005,kilicetal2015,kilicetal2018a}.

Three of the seven claimed pulsating ELM WDs show no RV variations \citep{sun&arras2018}. While these could be low-inclination binaries, a similar fraction (6/15) of non-pulsating ELM candidates studied by \citet{brownetal2016} have no RV variability, so there are too many such systems for low inclination to explain them all. Superficially, this suggests that not all ELM WDs are in binaries, contrary to theory. However, there is likely a problem with some ELM classifications. \citet{belletal2017,belletal2018} have outlined methods for excising interlopers from the class, such as high-amplitude $\delta$\,Sct stars with inaccurate spectroscopic $\log g$ values that masquerade as pulsating ELMs, and the large number of cooler subdwarfs (``sdAs'') with similar spectra. There remain only four bonafide pulsating ELM WD binaries, which are the four with RV variations. Among the non-pulsators, 85\% have now been shown to be RV variable, which, after accounting for false positives and the occasional system at low inclination, is consistent with all ELMs being in short-period binaries \citep{belletal2017}. This field is rapidly evolving, and both asteroseismology and binarity remain at its forefront.

\subsection{Hot subdwarfs}

The other class of objects populating the low-mass end of Fig.\,\ref{fig:wd} is the hot subdwarfs, also called sdB or sdO stars, depending on spectral type. Here we use the generalism `sdBs' to describe them all, though differences in character and evolution exist between the groups. These stars typically have masses between 0.30 and 0.49\,M$_{\odot}$ \citep{podsiadlowskietal2008}, and at least half are located in binaries \citep{maxtedetal2001,copperwheatetal2011}, though some sdBs may originate as the merger of two He WDs (\citealt{webbink1984}, cf. \citealt{schwab2018}). The masses depend on the details of the binary evolution pathway. Those forming via a common-envelope phase reside in short-period binaries (0.1 $< P <$ 10\,d), are more massive, and have either WD or low-mass main-sequence companions \citep{hanetal2003}. However, lower-mass sdBs can result from stable Roche-lobe overflow with main-sequence companions in wide binaries ($P$ $\lesssim$ 1300\,d). For a thorough review on sdBs, see \citet{heber2016}.

\nocite{ostensenetal2014a} 
\textit{Kepler} delivered remarkable light curves of pulsating sdBs, including some in eclipsing binaries where reflection, doppler beaming and gravitational lensing could be observed \citep{ostensenetal2010,bloemenetal2011}. \textit{Kepler} sdBs also include 18 pulsators \citep{teltingetal2014b}. Hotter sdBs can pulsate in p~modes with periods of a few minutes \citep{charpinetetal1996, kilkennyetal1997} and the cooler sdBs can pulsate in g~modes with periods of 0.8--2\,hr \citep{greenetal2003}. Those near the temperature boundary of 28,000\,K can pulsate in both p and g~modes simultaneously as hybrids \citep{schuhetal2006}. Advancements from asteroseismology of \textit{Kepler} sdBs included the detection of rotational splittings \citep{reedetal2014} and the discovery of equal spacings between the periods of g~modes \citep{ostensenetal2014b}, similar to those seen in the cores of red giants \citep{beddingetal2011}.

Although the instability strip of sdBs is not pure \citep{ostensenetal2011}, most sdBs are in binaries, and pulsation timing has seen success on these stars. \citet{teltingetal2014b} studied the sdB+WD binary KIC\,11558725 and solved its 10-d orbit in three ways: (i) with RVs; (ii) by the doppler beaming signal in the light curve; and (iii) by pulsation timing. Generally, the high oscillation frequencies of p~modes in sdBs compensate for their typically small semi-major axes, keeping pulsation timing viable. However, care must be taken because stochastic phase variations are common among sdBs \citep{ostensenetal2014a}. A pulsation-timing exoplanet was once claimed for V391\,Peg \citep{silvottietal2007}, but the signal from different oscillation modes now appears to be divergent, so the planet's existence has been downgraded to `putative' \citep{silvottietal2018}. Another example of intrinsic phase modulation comes from the detailed frequency analysis of the hybrid sdB pulsator KIC\,3527751, which showed phase and/or amplitude modulation in all strong oscillation modes \citep{zongetal2018}. While the amplitude modulation can potentially be used for mode identification \citep{giammicheleetal2018}, any pulsation timing effort will be complicated. The same modulation should be seen in all modes if the source is binary motion. Only one pulsation-timing exoplanet discovery remains plausible \citep{hermes2017}, which comes from the two-mode analysis of a $\delta$\,Sct star \citep{murphyetal2016c}, though it should be noted that the planet mass is close to the brown-dwarf mass range.

\begin{figure}
\centering
\includegraphics[width=0.99\linewidth]{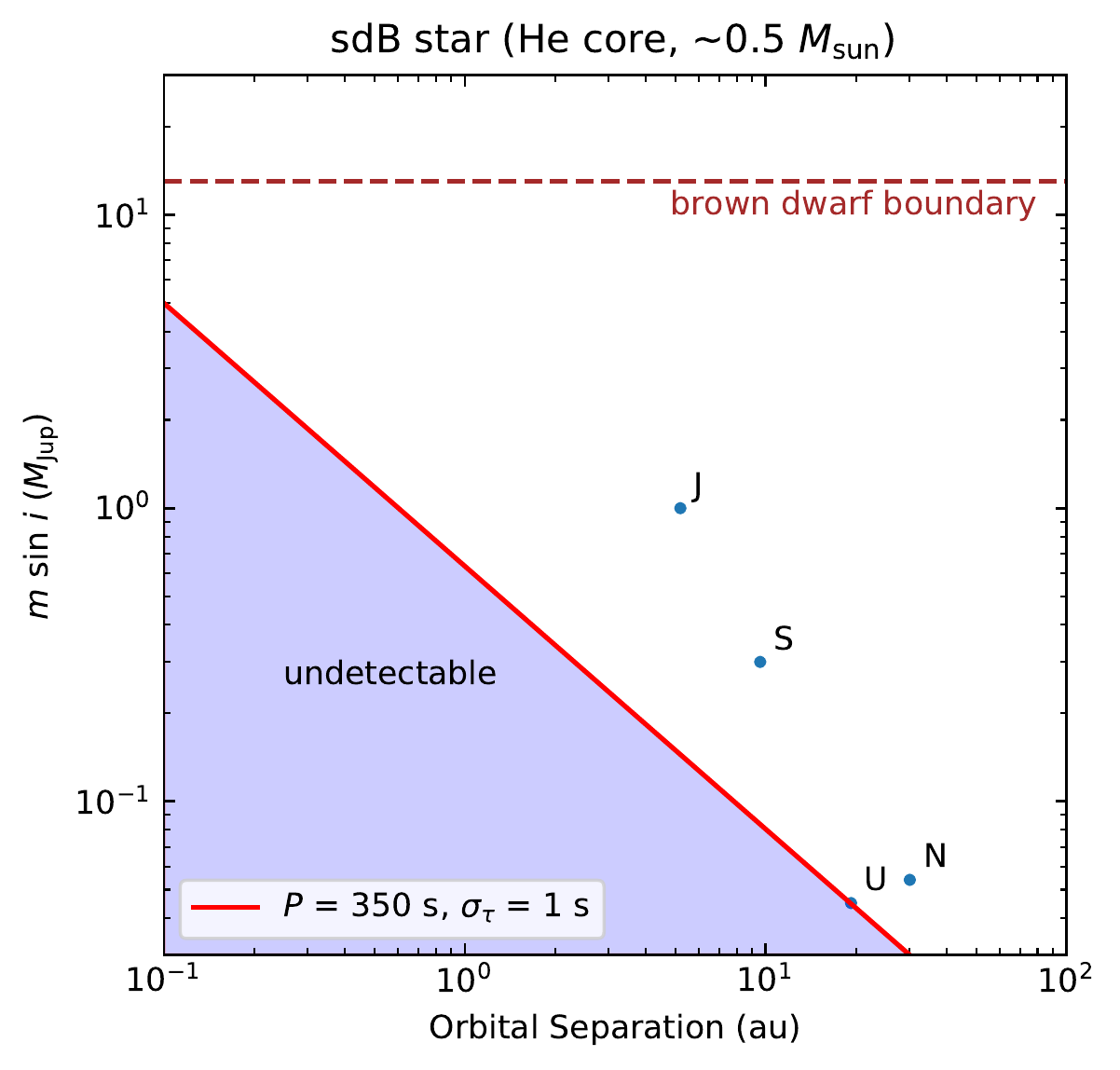}
\caption{Pulsation-timing detection limits on the mass of hypothetical exoplanets orbiting hot subdwarfs. The red line shows the limit for oscillation periods of 350\,s with intrinsic uncertainties on time-delay measurements of 1\,s. Modified from \citet{hermes2017}.}
\label{fig:sdB_planet}
\end{figure}

If an sdB star were to host an exoplanet in a wide orbit and pulsate in well-behaved p~modes, then pulsation timing could detect that planet if it resembles Jupiter or Saturn in mass and semi-major axis, while Uranus and Neptune would lie close to the detection limit (Fig.\,\ref{fig:sdB_planet}). Brown dwarfs could be detected at any plausible period. However, it is not yet established that subdwarf oscillations are coherent enough on timescales of Neptune's orbital period that this could work in practice. Instead, exoplanet searches have shifted focus to eclipse timings.

Eclipse timing variations have led to several claims of circumbinary companions to sdB stars, starting with two giant planets orbiting the pulsating-sdB prototype HW\,Vir \citep{leeetal2009}, but for every claim there is a conflicting analysis regarding the orbital stability or the persistence of the evidence between different datasets (see the review by \citealt{heber2016}). If real, these systems could shed light on the fate of exoplanets during the late stages of binary evolution.

A large population of sdBs was recently detected indirectly by pulsation timing of \textit{Kepler} $\delta$\,Sct stars \citep{murphyetal2018}. They found an excess of low-eccentricity companions at periods of several hundred days -- consistent with post-mass-transfer orbits. The companion masses were consistent with He-core remnants. For any given system it is not known whether the companion is an sdB or a more massive companion at low inclination, but statistically many will be sdBs and follow-up observations may well be rewarding.

Another phenomenon related to pulsation timing is helpful in the analysis of sdB orbits. The R\o{}mer delay describes the asymmetry of eclipse timings in a binary system due to light travel-time effects. In a perfectly circular orbit of an equal-mass binary, the secondary eclipses should arrive half way between primary eclipses, but a mass ratio smaller than unity causes an asymmetry in the path length travelled by light from the eclipsed star on its way to Earth. The effect is named for R\o{}mer's observation of the variation in eclipse arrival times of the Galilean moons passing behind Jupiter. It has been observed for sdB stars with M-dwarf companions and used to derive mass ratios \citep{barlowetal2012}, based on the theory from \citet{kaplan2010}.

However, one must be careful when using the R\o{}mer delay in this way, because similar delays in secondary eclipse arrival times arise naturally from orbital eccentricity. It was once assumed that the post-common-envelope orbits of sdB stars had zero eccentricity based on tidal circularisation theory \citep{zahn1977}, but other effects such as perturbing third bodies or spin-orbit misalignment before the common-envelop phase can induce eccentricity \citep{barlowetal2012}. \citet{preeceetal2018} have recently shown that tides are inefficient in circularising and synchronising sdB binaries. Unfortunately, it is difficult to measure the eccentricities of sdB orbits to the precision required to refine tidal circularisation theory ($e\lesssim10^{-5}$), but tests for tidally-induced spin--orbit synchronisation are much easier.

Half of the observed sdB stars lie in short-period binaries \citep{napiwotzkietal2004,copperwheatetal2011}, many of which eclipse or show phase-related effects like reflection (mutual heating), so their orbital periods are readily measured. Rotation periods can be determined spectroscopically if a radius is known, or asteroseismically if the star pulsates, to provide tests of tidal synchronisation. Intriguingly, of the many available test cases, only NY\,Vir demonstrates clear spin--orbit synchronisation \citep{charpinetetal2008}, whereas several systems rotate subsynchronously, i.e. with periods longer than the orbital period (B4, \citealt{pabloetal2011}; KIC\,2991403 and KIC\,11179657, \citealt{pabloetal2012b}; KIC\,10553698, \citealt{ostensenetal2014b}; KIC\,7668647, \citealt{teltingetal2014b};  J162256+473051, \citealt{schaffenrothetal2014}; PG1142-037, \citealt{reedetal2016}; and KIC\,7664467, \citealt{baranetal2016}). \citet{preeceetal2018} ascertained the cause, which is that in sdB stars the tidal synchronisation timescales are longer than the lifetime of their core-He-burning phase, because the orbital period is typically shorter than the convective turnover time-scale, damping the convective dissipation of tides. These developments in tidal synchronisation are yet another success story from studies of pulsating stars in binaries.

\subsection{White Dwarfs}

White dwarfs oscillate at very high frequencies, which could make them good clocks for detecting exoplanet companions by pulsation timing \citep{wingetetal2015,comptonetal2016}. However, intrinsic phase variability once again foils attempts to exploit this sensitivity \citep{hermesetal2013b,dalessioetal2013}. \citet{hermes2017} provided an overview of pulsation timing attempts on WDs, including the discovery and subsequent refutation of a possible substellar companion to the ZZ\,Ceti star GG\,66 (\citealt{mullallyetal2008}; cf. \citealt{dalessio2013}). An earlier case concerns the WD G29-38, whose phase variation observed in a single mode was once suspected to be due to an unseen companion \citep{wingetetal1990}, but turned out to be a false positive \citep{kleinmanetal1994}. Pulsating CO WDs in close binaries happen to be rare \citep{hermesetal2015}, and no orbits have thus far been determined by pulsation timing.

Fortunately, asteroseismic and spectroscopic modelling of WDs reach higher precision than is possible from binary analysis \citep{winget&kepler2008}. Surface gravities are routinely determined to better than $\pm0.1$\,dex, leading to precise masses and thence radii via the WD mass--radius relation \citep{fontaineetal2001}. This precision is further refined, from the opposite direction, when radii are calculable from \textit{Gaia}-derived luminosities and spectroscopic temperatures \citep{tremblayetal2017}. 

Describing the binary properties of WDs is extremely important for understanding the origin of Type Ia supernovae, which function as important cosmological distance indicators and contribute heavily to the chemical enrichment of galaxies \citep{maozetal2014}. While asteroseismology already offers clues to the WD properties, the most promising future avenue for studying their binarity perhaps lies with Gaia astrometry \citep{hollandsetal2018}.

\section{Tidally excited oscillations}
\label{sec:TEOs}

Thus far we have considered self-excited oscillations in stars that happen to be in binaries. We finish by considering the converse: oscillations excited by virtue of binary motion.

Tidally excited oscillations (TEOs) are gravity modes found in eccentric binaries as a result of tidal forcing (\citealt{fuller2017} provided an excellent overview). Their oscillation frequencies occur at exact integer multiples of the orbital frequency, particularly where these coincide with g-mode frequencies. Perhaps the best-studied case is KOI-54, in which TEOs are excited at 90 and 91 times the orbital frequency \citep{welshetal2011,fuller&lai2012,burkartetal2012,oleary&burkart2014}.

\begin{figure}
\centering
\includegraphics[width=0.99\linewidth]{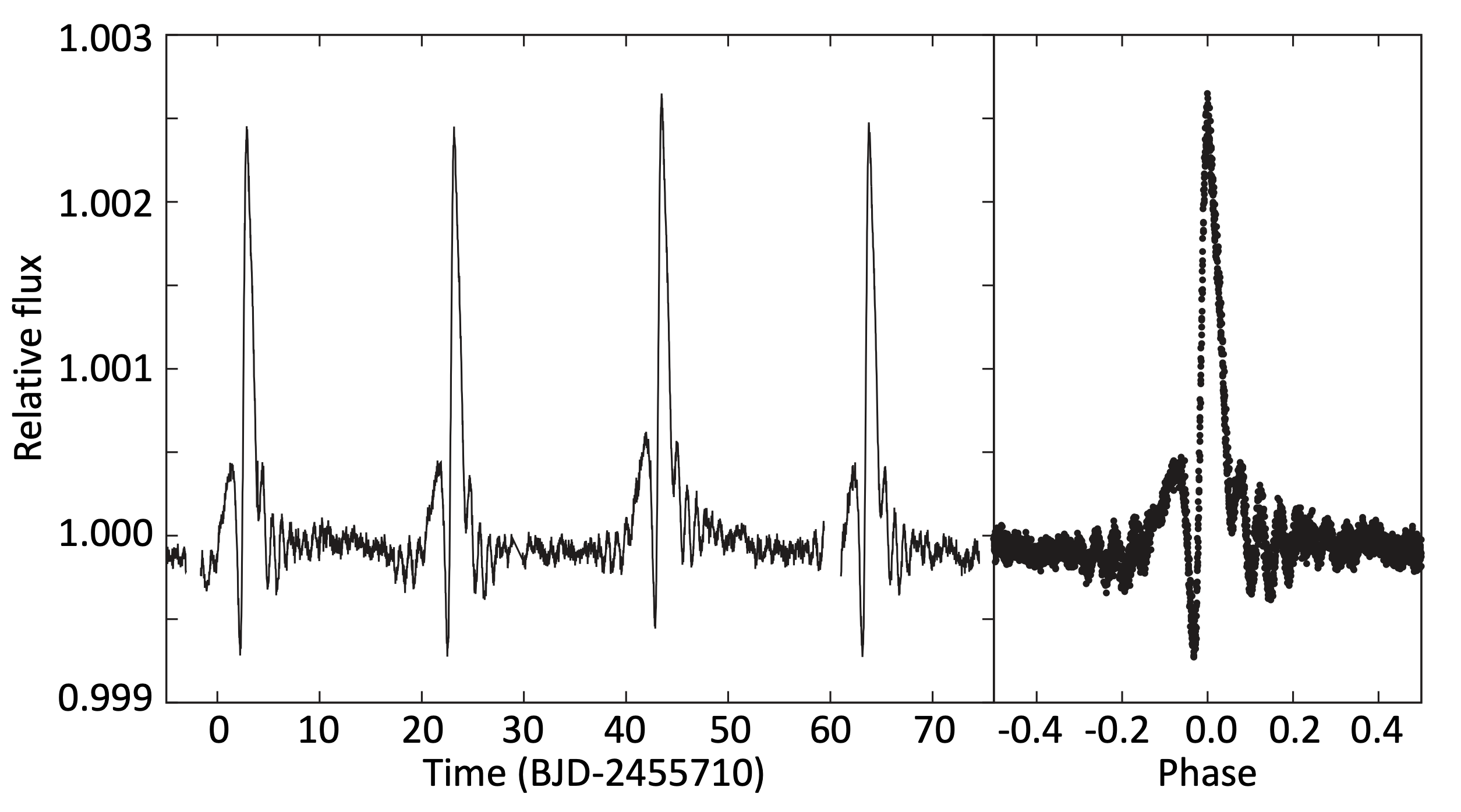}
\caption{Light curve of the heartbeat star KIC\,3749404. The dominant variations result from tidal deformation, reflection, mutual irradiation (heating) and Doppler beaming. Tidally excited oscillations, commensurate with the 20.3-d orbital period, are clear in the phased light curve ({\it right}). Modified from \citet{hambletonetal2016}.}
\label{fig:hb}
\end{figure}

TEOs are found in the light curves of `heartbeat' stars, so named because the periastron brightenings of these eccentric binaries resemble cardiograms (Fig.\,\ref{fig:hb}). The shape of the heartbeat depends on the inclination, eccentricity and longitude of periastron of the binary \citep{kumaretal1995}, with eclipses visible at higher inclinations. It is therefore possible to model the heartbeat to extract this information, which in turn provides crucial input to asteroseismic modelling. Many consecutive heartbeats can be studied to infer the rate of apsidal motion, offering observational tests of classical and general relativistic predictions \citep{hambletonetal2016}. The largest samples of heartbeat stars come from \textit{Kepler} data. \citet{thompsonetal2012} discovered 17 with intermediate-mass and main-sequence primaries, and \citet{becketal2014} later discovered 18 containing pulsating red giants.

Some high-amplitude TEOs are the result of resonant locking \citep{fuller2017,hambletonetal2018}, due to a feedback mechanism between the evolution of the star and its orbit. Importantly, resonantly locked modes accelerate the tidal dissipation rate, leading to more rapid circularisation of the orbit \citep{fulleretal2017}. Resonant locks may also occur among short-period ($P \lesssim 2$\,hr) WD binaries \citep{burkartetal2013}, enhancing the rate of tidal synchronization in WDs with orbits decaying due to gravitational radiation. This could affect the outcome (and time-scale) of the merger process, which in turn would enhance the rate of double-degenerate Type Ia supernovae. The progenitors of Type Ia supernovae remain a puzzle \citep{maozetal2014}, for which TEOs may well be a missing piece.

\section{Conclusions}
\label{sec:conclusions}

Asteroseismology has advanced the frontier of our knowledge of stellar structure and evolution. It has pierced the surface and illuminated the physical processes operating within. The rotation rates of the cores of stars are now measurable from the main sequence to the red giant branch, and beyond, as the cores are stripped of their envelopes and shine as one of a variety of classes of degenerate stellar remnants. The parameters of convective core overshooting, mixing lengths, and diffusion coefficients have transitioned from being arbitrarily assigned to observationally constrained. In the space-photometry era, the field has genuinely been revolutionised.

Binary and multiple systems have also benefitted heavily from the outstanding photometry. Light curves now have the precision to show such subtle effects as reflected light from an orbiting exoplanet, so Doppler beaming, tidal deformation and mutual irradiation are readily observable in hundreds of systems. Our understanding of binary evolution pathways across various orbital configurations continues to advance, as new populations such as extremely low mass white dwarfs and hot subdwarfs are observed in greater number.

At the superposition of these two advancing wavefronts sit the pulsating stars in binary systems, where the power of asteroseismology and orbital dynamics in providing masses and radii combine. Some stars oscillate by virtue of binary motion (tidally excited oscillations); some pulsators are occasionally found in binaries (the many \textit{Kepler} red giants); and in each case the information content is wonderfully enriched. There are now catalogues of binary stars discovered only by the orbital effects on the pulsations, which with present technology are undiscoverable by other means. Occasionally, two pulsators are found orbiting each other, allowing mass ratios to be determined without resorting to spectroscopy.

There remain problems to be solved in asteroseismology and binary analyses alike. For instance, mode identification remains a problem for $\delta$\,Sct stars, and the production channels and event rates for type Ia supernovae await refinement. While such problems may at first seem completely unrelated, they can be linked in unexpected ways. For our example above, developments in mode identification may allow us to determine masses and ages for $\delta$\,Sct stars, many of which have WD companions (found by pulsation timing) and may be future progenitors of type Ia supernovae. We can model whether a WD in these systems will reach the Chandresekhar limit following mass transfer or a merger, if the parameters of the components can be reliably determined. This may lead to refinements in the type Ia event rate. Thus advances in one area can spur progress in another. This is but one example of the interrelation between binarity and pulsation, many more of which have been given in this review. Despite abundant recent developments (or perhaps because of them), there are many problems like these to work on, and thankfully, many data to work with, as the new data pouring in from {\it TESS} and {\it Gaia} add to the archives of {\it Kepler} and {\it K2}. It can be difficult to decide what to work on first. I suggest starting with pulsating stars in binary systems.

\section*{Acknowledgments}

SJM thanks Jim Fuller, JJ Hermes, Timothy Van Reeth and Tim White for their helpful comments on this manuscript.

\bibliographystyle{phostproc}
\bibliography{sjm_bibliography.bib}

\begin{thebibliography}{186}
\providecommand{\natexlab}[1]{#1}

\bibitem[\protect\astroncite{{Addison} \emph{et~al.}}{2018}]{addisonetal2018}
{Addison}, B.~C., {Wang}, S., {Johnson}, M.~C., {Tinney}, C.~G., {Wright},
  D.~J., \emph{et~al.} 2018, \aj, 156, 197.

\bibitem[\protect\astroncite{{Aerts}}{2015}]{aerts2015}
{Aerts}, C. 2015, Astronomische Nachrichten, 336, 477.

\bibitem[\protect\astroncite{{Aerts} \emph{et~al.}}{2010}]{aertsetal2010}
{Aerts}, C., {Christensen-Dalsgaard}, J., \& {Kurtz}, D.~W. 2010,
  \emph{{Asteroseismology}} ({Springer-Verlag, Berlin}).

\bibitem[\protect\astroncite{{Aerts} \emph{et~al.}}{2018}]{aertsetal2018b}
{Aerts}, C., {Mathis}, S., \& {Rogers}, T. 2018, ArXiv e-prints.

\bibitem[\protect\astroncite{{Albrecht} \emph{et~al.}}{2014}]{albrechtetal2014}
{Albrecht}, S., {Winn}, J.~N., {Torres}, G., {Fabrycky}, D.~C., {Setiawan}, J.,
  \emph{et~al.} 2014, \apj, 785, 83.

\bibitem[\protect\astroncite{{Appourchaux}
  \emph{et~al.}}{2015}]{appourchauxetal2015}
{Appourchaux}, T., {Antia}, H.~M., {Ball}, W., {Creevey}, O., {Lebreton}, Y.,
  \emph{et~al.} 2015, \aap, 582, A25.

\bibitem[\protect\astroncite{{Appourchaux}
  \emph{et~al.}}{2012}]{appourchauxetal2012}
{Appourchaux}, T., {Chaplin}, W.~J., {Garc{\'{\i}}a}, R.~A., {Gruberbauer}, M.,
  {Verner}, G.~A., \emph{et~al.} 2012, \aap, 543, A54.

\bibitem[\protect\astroncite{{Armstrong}
  \emph{et~al.}}{2006}]{armstrongetal2006}
{Armstrong}, J.~T., {Mozurkewich}, D., {Hajian}, A.~R., {Johnston}, K.~J.,
  {Thessin}, R.~N., \emph{et~al.} 2006, \aj, 131, 2643.

\bibitem[\protect\astroncite{{Baglin} \emph{et~al.}}{2006}]{baglinetal2006}
{Baglin}, A., {Auvergne}, M., {Boisnard}, L., {Lam-Trong}, T., {Barge}, P.,
  \emph{et~al.} 2006, In \emph{36th COSPAR Scientific Assembly}, \emph{COSPAR,
  Plenary Meeting}, vol.~36, pp. 3749--+.

\bibitem[\protect\astroncite{{Ballot} \emph{et~al.}}{2011}]{ballotetal2011}
{Ballot}, J., {Barban}, C., \& {van't Veer-Menneret}, C. 2011, \aap, 531, A124.

\bibitem[\protect\astroncite{{Baran} \emph{et~al.}}{2016}]{baranetal2016}
{Baran}, A.~S., {Telting}, J.~H., {N{\'e}meth}, P., {{\O}stensen}, R.~H.,
  {Reed}, M.~D., \emph{et~al.} 2016, \aap, 585, A66.

\bibitem[\protect\astroncite{{Barlow} \emph{et~al.}}{2012}]{barlowetal2012}
{Barlow}, B.~N., {Wade}, R.~A., \& {Liss}, S.~E. 2012, \apj, 753, 101.

\bibitem[\protect\astroncite{{Barnes} \& {Moffett}}{1975}]{barnes&moffett1975}
{Barnes}, T.~G., III \& {Moffett}, T.~J. 1975, \aj, 80, 48.

\bibitem[\protect\astroncite{{Beck} \emph{et~al.}}{2014}]{becketal2014}
{Beck}, P.~G., {Hambleton}, K., {Vos}, J., {Kallinger}, T., {Bloemen}, S.,
  \emph{et~al.} 2014, \aap, 564, A36.

\bibitem[\protect\astroncite{{Bedding} \emph{et~al.}}{2004}]{beddingetal2004}
{Bedding}, T.~R., {Kjeldsen}, H., {Butler}, R.~P., {McCarthy}, C., {Marcy},
  G.~W., \emph{et~al.} 2004, \apj, 614, 380.

\bibitem[\protect\astroncite{{Bedding} \emph{et~al.}}{2011}]{beddingetal2011}
{Bedding}, T.~R., {Mosser}, B., {Huber}, D., {Montalb{\'a}n}, J., {Beck}, P.,
  \emph{et~al.} 2011, \nat, 471, 608.

\bibitem[\protect\astroncite{{Bell} \emph{et~al.}}{2017}]{belletal2017}
{Bell}, K.~J., {Gianninas}, A., {Hermes}, J.~J., {Winget}, D.~E., {Kilic}, M.,
  \emph{et~al.} 2017, \apj, 835, 180.

\bibitem[\protect\astroncite{{Bell} \emph{et~al.}}{2018}]{belletal2018}
{Bell}, K.~J., {Pelisoli}, I., {Kepler}, S.~O., {Brown}, W.~R., {Winget},
  D.~E., \emph{et~al.} 2018, \aap, 617, A6.

\bibitem[\protect\astroncite{{Bloemen} \emph{et~al.}}{2011}]{bloemenetal2011}
{Bloemen}, S., {Marsh}, T.~R., {{\O}stensen}, R.~H., {Charpinet}, S.,
  {Fontaine}, G., \emph{et~al.} 2011, \mnras, 410, 1787.

\bibitem[\protect\astroncite{{Boffin}}{2015}]{boffin2015}
{Boffin}, H.~M.~J. 2015, \aap, 575, L13.

\bibitem[\protect\astroncite{{Borucki} \emph{et~al.}}{2010}]{boruckietal2010}
{Borucki}, W.~J., {Koch}, D., {Basri}, G., {Batalha}, N., {Brown}, T.,
  \emph{et~al.} 2010, Science, 327, 977.

\bibitem[\protect\astroncite{{Bouchy} \& {Carrier}}{2002}]{bouchy&carrier2002}
{Bouchy}, F. \& {Carrier}, F. 2002, \aap, 390, 205.

\bibitem[\protect\astroncite{{Bowman} \emph{et~al.}}{2016}]{bowmanetal2016}
{Bowman}, D.~M., {Kurtz}, D.~W., {Breger}, M., {Murphy}, S.~J., \&
  {Holdsworth}, D.~L. 2016, \mnras, 460, 1970.

\bibitem[\protect\astroncite{{Breger}}{2000}]{breger2000b}
{Breger}, M. 2000, In \emph{IAU Colloq. 176: The Impact of Large-Scale Surveys
  on Pulsating Star Research}, edited by {L.~Szabados \& D.~Kurtz},
  \emph{Astronomical Society of the Pacific Conference Series}, vol. 203, pp.
  421--425 ({ASP Conference Series}).

\bibitem[\protect\astroncite{{Breger} \emph{et~al.}}{2002}]{bregeretal2002b}
{Breger}, M., {Pamyatnykh}, A.~A., {Zima}, W., {Garrido}, R., {Handler}, G.,
  \emph{et~al.} 2002, \mnras, 336, 249.

\bibitem[\protect\astroncite{{Brogaard} \emph{et~al.}}{2018}]{brogaardetal2018}
{Brogaard}, K., {Hansen}, C.~J., {Miglio}, A., {Slumstrup}, D., {Frandsen}, S.,
  \emph{et~al.} 2018, \mnras, 476, 3729.

\bibitem[\protect\astroncite{{Brown} \emph{et~al.}}{2016}]{brownetal2016}
{Brown}, W.~R., {Gianninas}, A., {Kilic}, M., {Kenyon}, S.~J., \& {Allende
  Prieto}, C. 2016, \apj, 818, 155.

\bibitem[\protect\astroncite{{Burkart} \emph{et~al.}}{2012}]{burkartetal2012}
{Burkart}, J., {Quataert}, E., {Arras}, P., \& {Weinberg}, N.~N. 2012, \mnras,
  421, 983.

\bibitem[\protect\astroncite{{Burkart} \emph{et~al.}}{2013}]{burkartetal2013}
{Burkart}, J., {Quataert}, E., {Arras}, P., \& {Weinberg}, N.~N. 2013, \mnras,
  433, 332.

\bibitem[\protect\astroncite{{Chaplin}
  \emph{et~al.}}{2011{\natexlab{a}}}]{chaplinetal2011}
{Chaplin}, W.~J., {Kjeldsen}, H., {Bedding}, T.~R., {Christensen-Dalsgaard},
  J., {Gilliland}, R.~L., \emph{et~al.} 2011{\natexlab{a}}, \apj, 732, 54.

\bibitem[\protect\astroncite{{Chaplin}
  \emph{et~al.}}{2011{\natexlab{b}}}]{chaplinetal2011a}
{Chaplin}, W.~J., {Kjeldsen}, H., {Christensen-Dalsgaard}, J., {Basu}, S.,
  {Miglio}, A., \emph{et~al.} 2011{\natexlab{b}}, Science, 332, 213.

\bibitem[\protect\astroncite{{Charbonneau} \&
  {Michaud}}{1991}]{charbonneau&michaud1991}
{Charbonneau}, P. \& {Michaud}, G. 1991, \apj, 370, 693.

\bibitem[\protect\astroncite{{Charpinet}
  \emph{et~al.}}{1996}]{charpinetetal1996}
{Charpinet}, S., {Fontaine}, G., {Brassard}, P., \& {Dorman}, B. 1996, \apjl,
  471, L103.

\bibitem[\protect\astroncite{{Charpinet}
  \emph{et~al.}}{2008}]{charpinetetal2008}
{Charpinet}, S., {Van Grootel}, V., {Reese}, D., {Fontaine}, G., {Green},
  E.~M., \emph{et~al.} 2008, \aap, 489, 377.

\bibitem[\protect\astroncite{{Colman} \emph{et~al.}}{2017}]{colmanetal2017}
{Colman}, I.~L., {Huber}, D., {Bedding}, T.~R., {Kuszlewicz}, J.~S., {Yu}, J.,
  \emph{et~al.} 2017, \mnras, 469, 3802.

\bibitem[\protect\astroncite{{Compton} \emph{et~al.}}{2016}]{comptonetal2016}
{Compton}, D.~L., {Bedding}, T.~R., {Murphy}, S.~J., \& {Stello}, D. 2016,
  \mnras, 461, 1943.

\bibitem[\protect\astroncite{{Copperwheat}
  \emph{et~al.}}{2011}]{copperwheatetal2011}
{Copperwheat}, C.~M., {Morales-Rueda}, L., {Marsh}, T.~R., {Maxted}, P.~F.~L.,
  \& {Heber}, U. 2011, \mnras, 415, 1381.

\bibitem[\protect\astroncite{{Cummings} \emph{et~al.}}{2018}]{cummingsetal2018}
{Cummings}, J.~D., {Kalirai}, J.~S., {Tremblay}, P.-E., {Ramirez-Ruiz}, E., \&
  {Choi}, J. 2018, \apj, 866, 21.

\bibitem[\protect\astroncite{{Dalessio} \emph{et~al.}}{2013}]{dalessioetal2013}
{Dalessio}, J., {Sullivan}, D.~J., {Provencal}, J.~L., {Shipman}, H.~L.,
  {Sullivan}, T., \emph{et~al.} 2013, \apj, 765, 5.

\bibitem[\protect\astroncite{{Dalessio}}{2013}]{dalessio2013}
{Dalessio}, J.~R. 2013, \emph{{Peculiar variations of white dwarf pulsation
  frequencies and maestro}}.
\newblock Ph.D. thesis, University of Delaware.

\bibitem[\protect\astroncite{{Davies} \emph{et~al.}}{2015}]{daviesetal2015}
{Davies}, G.~R., {Chaplin}, W.~J., {Farr}, W.~M., {Garc{\'\i}a}, R.~A., {Lund},
  M.~N., \emph{et~al.} 2015, \mnras, 446, 2959.

\bibitem[\protect\astroncite{{De Rosa} \emph{et~al.}}{2014}]{derosaetal2014}
{De Rosa}, R.~J., {Patience}, J., {Wilson}, P.~A., {Schneider}, A.,
  {Wiktorowicz}, S.~J., \emph{et~al.} 2014, \mnras, 437, 1216.

\bibitem[\protect\astroncite{{Duch{\^e}ne} \&
  {Kraus}}{2013}]{duchene&kraus2013}
{Duch{\^e}ne}, G. \& {Kraus}, A. 2013, \araa, 51, 269.

\bibitem[\protect\astroncite{{Dziembowski}}{1977}]{dziembowski1977}
{Dziembowski}, W. 1977, \actaa, 27, 203.

\bibitem[\protect\astroncite{{Fontaine} \emph{et~al.}}{2001}]{fontaineetal2001}
{Fontaine}, G., {Brassard}, P., \& {Bergeron}, P. 2001, \pasp, 113, 409.

\bibitem[\protect\astroncite{{Fuller}}{2017}]{fuller2017}
{Fuller}, J. 2017, \mnras, 472, 1538.

\bibitem[\protect\astroncite{{Fuller} \emph{et~al.}}{2017}]{fulleretal2017}
{Fuller}, J., {Hambleton}, K., {Shporer}, A., {Isaacson}, H., \& {Thompson}, S.
  2017, \mnras, 472, L25.

\bibitem[\protect\astroncite{{Fuller} \& {Lai}}{2012}]{fuller&lai2012}
{Fuller}, J. \& {Lai}, D. 2012, \mnras, 420, 3126.

\bibitem[\protect\astroncite{{Gamarova} \emph{et~al.}}{2003}]{gamarovaetal2003}
{Gamarova}, A.~Y., {Mkrtichian}, D.~E., {Rodriguez}, E., {Costa}, V., \&
  {Lopez-Gonzalez}, M.~J. 2003, In \emph{Interplay of Periodic, Cyclic and
  Stochastic Variability in Selected Areas of the H-R Diagram}, edited by
  C.~{Sterken}, \emph{Astronomical Society of the Pacific Conference Series},
  vol. 292, p. 369.

\bibitem[\protect\astroncite{{Gaulme} \& {Guzik}}{2014}]{guzik&gaulme2014}
{Gaulme}, P. \& {Guzik}, J.~A. 2014, In \emph{Precision Asteroseismology},
  edited by J.~A. {Guzik}, W.~J. {Chaplin}, G.~{Handler}, \& A.~{Pigulski},
  \emph{IAU Symposium}, vol. 301, pp. 413--414.

\bibitem[\protect\astroncite{{Gaulme} \emph{et~al.}}{2016}]{gaulmeetal2016}
{Gaulme}, P., {McKeever}, J., {Jackiewicz}, J., {Rawls}, M.~L., {Corsaro}, E.,
  \emph{et~al.} 2016, \apj, 832, 121.

\bibitem[\protect\astroncite{{Gebran} \emph{et~al.}}{2010}]{gebranetal2010}
{Gebran}, M., {Vick}, M., {Monier}, R., \& {Fossati}, L. 2010, \aap, 523, A71.

\bibitem[\protect\astroncite{{Giammichele}
  \emph{et~al.}}{2018}]{giammicheleetal2018}
{Giammichele}, N., {Charpinet}, S., {Fontaine}, G., {Brassard}, P., {Green},
  E.~M., \emph{et~al.} 2018, \nat, 554, 73.

\bibitem[\protect\astroncite{{Gilliland}
  \emph{et~al.}}{2010{\natexlab{a}}}]{gillilandetal2010a}
{Gilliland}, R.~L., {Brown}, T.~M., {Christensen-Dalsgaard}, J., {Kjeldsen},
  H., {Aerts}, C., \emph{et~al.} 2010{\natexlab{a}}, \pasp, 122, 131.

\bibitem[\protect\astroncite{{Gilliland}
  \emph{et~al.}}{2010{\natexlab{b}}}]{gillilandetal2010b}
{Gilliland}, R.~L., {Jenkins}, J.~M., {Borucki}, W.~J., {Bryson}, S.~T.,
  {Caldwell}, D.~A., \emph{et~al.} 2010{\natexlab{b}}, \apjl, 713, L160.

\bibitem[\protect\astroncite{{Girardi}}{2016}]{girardi2016}
{Girardi}, L. 2016, Astronomische Nachrichten, 337, 871.

\bibitem[\protect\astroncite{{Gizon} \& {Solanki}}{2003}]{gizon&solanki2003}
{Gizon}, L. \& {Solanki}, S.~K. 2003, \apj, 589, 1009.

\bibitem[\protect\astroncite{{Green} \emph{et~al.}}{2003}]{greenetal2003}
{Green}, E.~M., {Fontaine}, G., {Reed}, M.~D., {Callerame}, K., {Seitenzahl},
  I.~R., \emph{et~al.} 2003, \apjl, 583, L31.

\bibitem[\protect\astroncite{{Guggenberger}
  \emph{et~al.}}{2016}]{guggenbergeretal2016}
{Guggenberger}, E., {Hekker}, S., {Basu}, S., \& {Bellinger}, E. 2016, \mnras,
  460, 4277.

\bibitem[\protect\astroncite{{Guszejnov}
  \emph{et~al.}}{2017}]{guszejnovetal2017}
{Guszejnov}, D., {Hopkins}, P.~F., \& {Krumholz}, M.~R. 2017, \mnras, 468,
  4093.

\bibitem[\protect\astroncite{{Hambleton}
  \emph{et~al.}}{2018}]{hambletonetal2018}
{Hambleton}, K., {Fuller}, J., {Thompson}, S., {Pr{\v s}a}, A., {Kurtz}, D.~W.,
  \emph{et~al.} 2018, \mnras, 473, 5165.

\bibitem[\protect\astroncite{{Hambleton}
  \emph{et~al.}}{2016}]{hambletonetal2016}
{Hambleton}, K., {Kurtz}, D.~W., {Pr{\v s}a}, A., {Quinn}, S.~N., {Fuller}, J.,
  \emph{et~al.} 2016, \mnras, 463, 1199.

\bibitem[\protect\astroncite{{Han} \emph{et~al.}}{2003}]{hanetal2003}
{Han}, Z., {Podsiadlowski}, P., {Maxted}, P.~F.~L., \& {Marsh}, T.~R. 2003,
  \mnras, 341, 669.

\bibitem[\protect\astroncite{{Handberg} \emph{et~al.}}{2017}]{handbergetal2017}
{Handberg}, R., {Brogaard}, K., {Miglio}, A., {Bossini}, D., {Elsworth}, Y.,
  \emph{et~al.} 2017, \mnras, 472, 979.

\bibitem[\protect\astroncite{{Heber}}{2009}]{heber2009}
{Heber}, U. 2009, \araa, 47, 211.

\bibitem[\protect\astroncite{{Heber}}{2016}]{heber2016}
{Heber}, U. 2016, \pasp, 128, 082001.

\bibitem[\protect\astroncite{{He{\l}miniak}
  \emph{et~al.}}{2017}]{helminiaketal2017}
{He{\l}miniak}, K.~G., {Ukita}, N., {Kambe}, E., {Koz{\l}owski}, S.~K.,
  {Paw{\l}aszek}, R., \emph{et~al.} 2017, \aap, 602, A30.

\bibitem[\protect\astroncite{{Hermes}}{2017}]{hermes2017}
{Hermes}, J.~J. 2017, \emph{{Timing by Stellar Pulsations as an Exoplanet
  Discovery Method}}, p.~6.

\bibitem[\protect\astroncite{{Hermes} \emph{et~al.}}{2015}]{hermesetal2015}
{Hermes}, J.~J., {G{\"a}nsicke}, B.~T., {Bischoff-Kim}, A., {Kawaler}, S.~D.,
  {Fuchs}, J.~T., \emph{et~al.} 2015, \mnras, 451, 1701.

\bibitem[\protect\astroncite{{Hermes}
  \emph{et~al.}}{2013{\natexlab{a}}}]{hermesetal2013d}
{Hermes}, J.~J., {Montgomery}, M.~H., {Gianninas}, A., {Winget}, D.~E.,
  {Brown}, W.~R., \emph{et~al.} 2013{\natexlab{a}}, \mnras, 436, 3573.

\bibitem[\protect\astroncite{{Hermes}
  \emph{et~al.}}{2013{\natexlab{b}}}]{hermesetal2013b}
{Hermes}, J.~J., {Montgomery}, M.~H., {Mullally}, F., {Winget}, D.~E., \&
  {Bischoff-Kim}, A. 2013{\natexlab{b}}, \apj, 766, 42.

\bibitem[\protect\astroncite{{Hermes}
  \emph{et~al.}}{2013{\natexlab{c}}}]{hermesetal2013a}
{Hermes}, J.~J., {Montgomery}, M.~H., {Winget}, D.~E., {Brown}, W.~R.,
  {Gianninas}, A., \emph{et~al.} 2013{\natexlab{c}}, \apj, 765, 102.

\bibitem[\protect\astroncite{{Hill} \&
  {Landstreet}}{1993}]{hill&landstreet1993}
{Hill}, G.~M. \& {Landstreet}, J.~D. 1993, \aap, 276, 142.

\bibitem[\protect\astroncite{{Hollands} \emph{et~al.}}{2018}]{hollandsetal2018}
{Hollands}, M.~A., {Tremblay}, P.-E., {G{\"a}nsicke}, B.~T., {Gentile-Fusillo},
  N.~P., \& {Toonen}, S. 2018, \mnras, 480, 3942.

\bibitem[\protect\astroncite{{Huang} \emph{et~al.}}{2018}]{huangetal2019}
{Huang}, C.~X., {Burt}, J., {Vanderburg}, A., {G{\"u}nther}, M.~N., {Shporer},
  A., \emph{et~al.} 2018, ArXiv e-prints.

\bibitem[\protect\astroncite{{Huber} \emph{et~al.}}{2013}]{huberetal2013}
{Huber}, D., {Carter}, J.~A., {Barbieri}, M., {Miglio}, A., {Deck}, K.~M.,
  \emph{et~al.} 2013, Science, 342, 331.

\bibitem[\protect\astroncite{{Jacoby} \emph{et~al.}}{2003}]{jacobyetal2003}
{Jacoby}, B.~A., {Bailes}, M., {van Kerkwijk}, M.~H., {Ord}, S., {Hotan}, A.,
  \emph{et~al.} 2003, \apjl, 599, L99.

\bibitem[\protect\astroncite{{Jacoby} \emph{et~al.}}{2005}]{jacobyetal2005}
{Jacoby}, B.~A., {Hotan}, A., {Bailes}, M., {Ord}, S., \& {Kulkarni}, S.~R.
  2005, \apjl, 629, L113.

\bibitem[\protect\astroncite{{Johnston} \emph{et~al.}}{2019}]{johnstonetal2019}
{Johnston}, C., {Tkachenko}, A., {Aerts}, C., {Molenberghs}, G., {Bowman},
  D.~M., \emph{et~al.} 2019, \mnras, 482, 1231.

\bibitem[\protect\astroncite{{Joyce} \& {Chaboyer}}{2018}]{joyce&chaboyer2018}
{Joyce}, M. \& {Chaboyer}, B. 2018, \apj, 864, 99.

\bibitem[\protect\astroncite{{Kallinger}
  \emph{et~al.}}{2018}]{kallingeretal2018}
{Kallinger}, T., {Beck}, P.~G., {Stello}, D., \& {Garcia}, R.~A. 2018, \aap,
  616, A104.

\bibitem[\protect\astroncite{{Kaplan}}{2010}]{kaplan2010}
{Kaplan}, D.~L. 2010, \apjl, 717, L108.

\bibitem[\protect\astroncite{{Keen} \emph{et~al.}}{2015}]{keenetal2015}
{Keen}, M.~A., {Bedding}, T.~R., {Murphy}, S.~J., {Schmid}, V.~S., {Aerts}, C.,
  \emph{et~al.} 2015, \mnras, 454, 1792.

\bibitem[\protect\astroncite{{Kilic} \emph{et~al.}}{2018}]{kilicetal2018a}
{Kilic}, M., {Hermes}, J.~J., {C{\'o}rsico}, A.~H., {Kosakowski}, A., {Brown},
  W.~R., \emph{et~al.} 2018, \mnras, 479, 1267.

\bibitem[\protect\astroncite{{Kilic} \emph{et~al.}}{2015}]{kilicetal2015}
{Kilic}, M., {Hermes}, J.~J., {Gianninas}, A., \& {Brown}, W.~R. 2015, \mnras,
  446, L26.

\bibitem[\protect\astroncite{{Kilkenny} \emph{et~al.}}{1997}]{kilkennyetal1997}
{Kilkenny}, D., {Koen}, C., {O'Donoghue}, D., \& {Stobie}, R.~S. 1997, \mnras,
  285, 640.

\bibitem[\protect\astroncite{{Kirk} \emph{et~al.}}{2016}]{kirketal2016}
{Kirk}, B., {Conroy}, K., {Pr{\v s}a}, A., {Abdul-Masih}, M., {Kochoska}, A.,
  \emph{et~al.} 2016, \aj, 151, 68.

\bibitem[\protect\astroncite{{Kjeldsen} \&
  {Bedding}}{1995}]{kjeldsen&bedding1995}
{Kjeldsen}, H. \& {Bedding}, T.~R. 1995, \aap, 293, 87.

\bibitem[\protect\astroncite{{Kjeldsen} \emph{et~al.}}{2005}]{kjeldsenetal2005}
{Kjeldsen}, H., {Bedding}, T.~R., {Butler}, R.~P., {Christensen-Dalsgaard}, J.,
  {Kiss}, L.~L., \emph{et~al.} 2005, \apj, 635, 1281.

\bibitem[\protect\astroncite{{Kleinman} \emph{et~al.}}{1994}]{kleinmanetal1994}
{Kleinman}, S.~J., {Nather}, R.~E., {Winget}, D.~E., {Clemens}, J.~C.,
  {Bradley}, P.~A., \emph{et~al.} 1994, \apj, 436, 875.

\bibitem[\protect\astroncite{{Koch} \emph{et~al.}}{2010}]{kochetal2010}
{Koch}, D.~G., {Borucki}, W.~J., {Basri}, G., {Batalha}, N.~M., {Brown}, T.~M.,
  \emph{et~al.} 2010, \apjl, 713, L79.

\bibitem[\protect\astroncite{{Kratter} \&
  {Matzner}}{2006}]{kratter&matzner2006}
{Kratter}, K.~M. \& {Matzner}, C.~D. 2006, \mnras, 373, 1563.

\bibitem[\protect\astroncite{{Kratter} \emph{et~al.}}{2010}]{kratteretal2010a}
{Kratter}, K.~M., {Matzner}, C.~D., {Krumholz}, M.~R., \& {Klein}, R.~I. 2010,
  \apj, 708, 1585.

\bibitem[\protect\astroncite{{Kumar} \emph{et~al.}}{1995}]{kumaretal1995}
{Kumar}, P., {Ao}, C.~O., \& {Quataert}, E.~J. 1995, \apj, 449, 294.

\bibitem[\protect\astroncite{{Kurtz} \emph{et~al.}}{2014}]{kurtzetal2014}
{Kurtz}, D.~W., {Saio}, H., {Takata}, M., {Shibahashi}, H., {Murphy}, S.~J.,
  \emph{et~al.} 2014, \mnras, 444, 102.

\bibitem[\protect\astroncite{{Lampens} \emph{et~al.}}{2017}]{lampensetal2017}
{Lampens}, P., {Fr{\'e}mat}, Y., {Vermeylen}, L., {S{\'o}dor}, {\'A}.,
  {Skarka}, M., \emph{et~al.} 2017, ArXiv e-prints.

\bibitem[\protect\astroncite{{Ledoux}}{1951}]{ledoux1951}
{Ledoux}, P. 1951, \apj, 114, 373.

\bibitem[\protect\astroncite{{Lee} \emph{et~al.}}{2009}]{leeetal2009}
{Lee}, J.~W., {Kim}, S.-L., {Kim}, C.-H., {Koch}, R.~H., {Lee}, C.-U.,
  \emph{et~al.} 2009, \aj, 137, 3181.

\bibitem[\protect\astroncite{{Li}
  \emph{et~al.}}{2018{\natexlab{a}}}]{glietal2018}
{Li}, G., {Bedding}, T.~R., {Murphy}, S.~J., {Van Reeth}, T., {Antoci}, V.,
  \emph{et~al.} 2018{\natexlab{a}}, ArXiv e-prints.

\bibitem[\protect\astroncite{{Li}
  \emph{et~al.}}{2018{\natexlab{b}}}]{tlietal2018}
{Li}, T., {Bedding}, T.~R., {Huber}, D., {Ball}, W.~H., {Stello}, D.,
  \emph{et~al.} 2018{\natexlab{b}}, \mnras, 475, 981.

\bibitem[\protect\astroncite{{Li}
  \emph{et~al.}}{2018{\natexlab{c}}}]{ylietal2018}
{Li}, Y., {Bedding}, T.~R., {Li}, T., {Bi}, S., {Murphy}, S.~J., \emph{et~al.}
  2018{\natexlab{c}}, \mnras, 476, 470.

\bibitem[\protect\astroncite{{Liakos} \&
  {Niarchos}}{2017}]{liakos&niarchos2017}
{Liakos}, A. \& {Niarchos}, P. 2017, \mnras, 465, 1181.

\bibitem[\protect\astroncite{{Ligni{\`e}res} \&
  {Georgeot}}{2009}]{lignieres&georgeot2009}
{Ligni{\`e}res}, F. \& {Georgeot}, B. 2009, \aap, 500, 1173.

\bibitem[\protect\astroncite{{Lund} \emph{et~al.}}{2017}]{lundetal2017}
{Lund}, M.~N., {Silva Aguirre}, V., {Davies}, G.~R., {Chaplin}, W.~J.,
  {Christensen-Dalsgaard}, J., \emph{et~al.} 2017, \apj, 835, 172.

\bibitem[\protect\astroncite{{Maoz} \emph{et~al.}}{2014}]{maozetal2014}
{Maoz}, D., {Mannucci}, F., \& {Nelemans}, G. 2014, \araa, 52, 107.

\bibitem[\protect\astroncite{{Marsh} \emph{et~al.}}{1995}]{marshetal1995}
{Marsh}, T.~R., {Dhillon}, V.~S., \& {Duck}, S.~R. 1995, \mnras, 275, 828.

\bibitem[\protect\astroncite{{Maxted} \emph{et~al.}}{2001}]{maxtedetal2001}
{Maxted}, P.~F.~L., {Heber}, U., {Marsh}, T.~R., \& {North}, R.~C. 2001,
  \mnras, 326, 1391.

\bibitem[\protect\astroncite{{Metcalfe} \emph{et~al.}}{2012}]{metcalfeetal2012}
{Metcalfe}, T.~S., {Chaplin}, W.~J., {Appourchaux}, T., {Garc{\'\i}a}, R.~A.,
  {Basu}, S., \emph{et~al.} 2012, \apj, 748, L10.

\bibitem[\protect\astroncite{{Metcalfe} \emph{et~al.}}{2015}]{metcalfeetal2015}
{Metcalfe}, T.~S., {Creevey}, O.~L., \& {Davies}, G.~R. 2015, \apj, 811, L37.

\bibitem[\protect\astroncite{{Miglio} \emph{et~al.}}{2014}]{miglioetal2014}
{Miglio}, A., {Chaplin}, W.~J., {Farmer}, R., {Kolb}, U., {Girardi}, L.,
  \emph{et~al.} 2014, \apjl, 784, L3.

\bibitem[\protect\astroncite{{Mkrtichian}
  \emph{et~al.}}{2004}]{mkrtichianetal2004}
{Mkrtichian}, D.~E., {Kusakin}, A.~V., {Rodriguez}, E., {Gamarova}, A.~Y.,
  {Kim}, C., \emph{et~al.} 2004, \aap, 419, 1015.

\bibitem[\protect\astroncite{{Moe} \& {Di Stefano}}{2017}]{moe&distefano2017}
{Moe}, M. \& {Di Stefano}, R. 2017, \apjs, 230, 15.

\bibitem[\protect\astroncite{{Mullally} \emph{et~al.}}{2008}]{mullallyetal2008}
{Mullally}, F., {Winget}, D.~E., {Degennaro}, S., {Jeffery}, E., {Thompson},
  S.~E., \emph{et~al.} 2008, \apj, 676, 573.

\bibitem[\protect\astroncite{{Murphy}
  \emph{et~al.}}{2016{\natexlab{a}}}]{murphyetal2016c}
{Murphy}, S.~J., {Bedding}, T.~R., \& {Shibahashi}, H. 2016{\natexlab{a}},
  \apjl, 827, L17.

\bibitem[\protect\astroncite{{Murphy} \emph{et~al.}}{2014}]{murphyetal2014}
{Murphy}, S.~J., {Bedding}, T.~R., {Shibahashi}, H., {Kurtz}, D.~W., \&
  {Kjeldsen}, H. 2014, \mnras, 441, 2515.

\bibitem[\protect\astroncite{{Murphy}
  \emph{et~al.}}{2016{\natexlab{b}}}]{murphyetal2016a}
{Murphy}, S.~J., {Fossati}, L., {Bedding}, T.~R., {Saio}, H., {Kurtz}, D.~W.,
  \emph{et~al.} 2016{\natexlab{b}}, \mnras, 459, 1201.

\bibitem[\protect\astroncite{{Murphy} \emph{et~al.}}{2018}]{murphyetal2018}
{Murphy}, S.~J., {Moe}, M., {Kurtz}, D.~W., {Bedding}, T.~R., {Shibahashi}, H.,
  \emph{et~al.} 2018, \mnras, 474, 4322.

\bibitem[\protect\astroncite{{Murphy} \&
  {Shibahashi}}{2015}]{murphy&shibahashi2015}
{Murphy}, S.~J. \& {Shibahashi}, H. 2015, \mnras, 450, 4475.

\bibitem[\protect\astroncite{{Murphy}
  \emph{et~al.}}{2016{\natexlab{c}}}]{murphyetal2016b}
{Murphy}, S.~J., {Shibahashi}, H., \& {Bedding}, T.~R. 2016{\natexlab{c}},
  \mnras, 461, 4215.

\bibitem[\protect\astroncite{{Napiwotzki}
  \emph{et~al.}}{2004}]{napiwotzkietal2004}
{Napiwotzki}, R., {Karl}, C.~A., {Lisker}, T., {Heber}, U., {Christlieb}, N.,
  \emph{et~al.} 2004, \apss, 291, 321.

\bibitem[\protect\astroncite{{O'Leary} \& {Burkart}}{2014}]{oleary&burkart2014}
{O'Leary}, R.~M. \& {Burkart}, J. 2014, \mnras, 440, 3036.

\bibitem[\protect\astroncite{{{\O}stensen}
  \emph{et~al.}}{2010}]{ostensenetal2010}
{{\O}stensen}, R.~H., {Green}, E.~M., {Bloemen}, S., {Marsh}, T.~R., {Laird},
  J.~B., \emph{et~al.} 2010, \mnras, 408, L51.

\bibitem[\protect\astroncite{{{\O}stensen}
  \emph{et~al.}}{2014{\natexlab{a}}}]{ostensenetal2014a}
{{\O}stensen}, R.~H., {Reed}, M.~D., {Baran}, A.~S., \& {Telting}, J.~H.
  2014{\natexlab{a}}, \aap, 564, L14.

\bibitem[\protect\astroncite{{{\O}stensen}
  \emph{et~al.}}{2011}]{ostensenetal2011}
{{\O}stensen}, R.~H., {Silvotti}, R., {Charpinet}, S., {Oreiro}, R., {Bloemen},
  S., \emph{et~al.} 2011, \mnras, 414, 2860.

\bibitem[\protect\astroncite{{{\O}stensen}
  \emph{et~al.}}{2014{\natexlab{b}}}]{ostensenetal2014b}
{{\O}stensen}, R.~H., {Telting}, J.~H., {Reed}, M.~D., {Baran}, A.~S.,
  {Nemeth}, P., \emph{et~al.} 2014{\natexlab{b}}, \aap, 569, A15.

\bibitem[\protect\astroncite{{Ouazzani} \emph{et~al.}}{2017}]{ouazzanietal2017}
{Ouazzani}, R.-M., {Salmon}, S.~J.~A.~J., {Antoci}, V., {Bedding}, T.~R.,
  {Murphy}, S.~J., \emph{et~al.} 2017, \mnras, 465, 2294.

\bibitem[\protect\astroncite{{Pablo} \emph{et~al.}}{2011}]{pabloetal2011}
{Pablo}, H., {Kawaler}, S.~D., \& {Green}, E.~M. 2011, \apjl, 740, L47.

\bibitem[\protect\astroncite{{Pablo} \emph{et~al.}}{2012}]{pabloetal2012b}
{Pablo}, H., {Kawaler}, S.~D., {Reed}, M.~D., {Bloemen}, S., {Charpinet}, S.,
  \emph{et~al.} 2012, \mnras, 422, 1343.

\bibitem[\protect\astroncite{{Pietrukowicz}
  \emph{et~al.}}{2017}]{pietrukowiczetal2017}
{Pietrukowicz}, P., {Dziembowski}, W.~A., {Latour}, M., {Angeloni}, R.,
  {Poleski}, R., \emph{et~al.} 2017, Nature Astronomy, 1, 0166.

\bibitem[\protect\astroncite{{Podsiadlowski}
  \emph{et~al.}}{2008}]{podsiadlowskietal2008}
{Podsiadlowski}, P., {Han}, Z., {Lynas-Gray}, A.~E., \& {Brown}, D. 2008, In
  \emph{Hot Subdwarf Stars and Related Objects}, edited by U.~{Heber}, C.~S.
  {Jeffery}, \& R.~{Napiwotzki}, \emph{Astronomical Society of the Pacific
  Conference Series}, vol. 392, p.~15.

\bibitem[\protect\astroncite{{Preece} \emph{et~al.}}{2018}]{preeceetal2018}
{Preece}, H.~P., {Tout}, C.~A., \& {Jeffery}, C.~S. 2018, \mnras, 481, 715.

\bibitem[\protect\astroncite{{Rawls}}{2016}]{rawls2016}
{Rawls}, M.~L. 2016, \emph{{Red Giants in Eclipsing Binaries as a Benchmark for
  Asteroseismology}}.
\newblock Ph.D. thesis, New Mexico State University.

\bibitem[\protect\astroncite{{Reed} \emph{et~al.}}{2016}]{reedetal2016}
{Reed}, M.~D., {Baran}, A.~S., {{\O}stensen}, R.~H., {Telting}, J.~H., {Kern},
  J.~W., \emph{et~al.} 2016, \mnras, 458, 1417.

\bibitem[\protect\astroncite{{Reed} \emph{et~al.}}{2014}]{reedetal2014}
{Reed}, M.~D., {Foster}, H., {Telting}, J.~H., {{\O}stensen}, R.~H., {Farris},
  L.~H., \emph{et~al.} 2014, \mnras, 440, 3809.

\bibitem[\protect\astroncite{{Reese} \emph{et~al.}}{2013}]{reeseetal2013}
{Reese}, D.~R., {Prat}, V., {Barban}, C., {van 't Veer-Menneret}, C., \&
  {MacGregor}, K.~B. 2013, \aap, 550, A77.

\bibitem[\protect\astroncite{{Ricker} \emph{et~al.}}{2015}]{rickeretal2015}
{Ricker}, G.~R., {Winn}, J.~N., {Vanderspek}, R., {Latham}, D.~W., {Bakos},
  G.~{\'A}., \emph{et~al.} 2015, Journal of Astronomical Telescopes,
  Instruments, and Systems, 1, 014003.

\bibitem[\protect\astroncite{{Rizzuto} \emph{et~al.}}{2013}]{rizzutoetal2013}
{Rizzuto}, A.~C., {Ireland}, M.~J., {Robertson}, J.~G., {Kok}, Y., {Tuthill},
  P.~G., \emph{et~al.} 2013, \mnras, 436, 1694.

\bibitem[\protect\astroncite{{Rodrigues}
  \emph{et~al.}}{2017}]{rodriguesetal2017}
{Rodrigues}, T.~S., {Bossini}, D., {Miglio}, A., {Girardi}, L.,
  {Montalb{\'a}n}, J., \emph{et~al.} 2017, \mnras, 467, 1433.

\bibitem[\protect\astroncite{{Royer} \emph{et~al.}}{2007}]{royeretal2007}
{Royer}, F., {Zorec}, J., \& {G{\'o}mez}, A.~E. 2007, \aap, 463, 671.

\bibitem[\protect\astroncite{{Saio}}{1981}]{saio1981}
{Saio}, H. 1981, \apj, 244, 299.

\bibitem[\protect\astroncite{{Saio} \emph{et~al.}}{2018}]{saioetal2018a}
{Saio}, H., {Kurtz}, D.~W., {Murphy}, S.~J., {Antoci}, V.~L., \& {Lee}, U.
  2018, \mnras, 474, 2774.

\bibitem[\protect\astroncite{{Saio} \emph{et~al.}}{2015}]{saioetal2015}
{Saio}, H., {Kurtz}, D.~W., {Takata}, M., {Shibahashi}, H., {Murphy}, S.~J.,
  \emph{et~al.} 2015, \mnras, 447, 3264.

\bibitem[\protect\astroncite{{Schaffenroth}
  \emph{et~al.}}{2014}]{schaffenrothetal2014}
{Schaffenroth}, V., {Geier}, S., {Heber}, U., {Kupfer}, T., {Ziegerer}, E.,
  \emph{et~al.} 2014, \aap, 564, A98.

\bibitem[\protect\astroncite{{Schmid} \& {Aerts}}{2016}]{schmid&aerts2016}
{Schmid}, V.~S. \& {Aerts}, C. 2016, \aap, 592, A116.

\bibitem[\protect\astroncite{{Schmid} \emph{et~al.}}{2015}]{schmidetal2015}
{Schmid}, V.~S., {Tkachenko}, A., {Aerts}, C., {Degroote}, P., {Bloemen}, S.,
  \emph{et~al.} 2015, \aap, 584, A35.

\bibitem[\protect\astroncite{{Schuh} \emph{et~al.}}{2006}]{schuhetal2006}
{Schuh}, S., {Huber}, J., {Dreizler}, S., {Heber}, U., {O'Toole}, S.~J.,
  \emph{et~al.} 2006, \aap, 445, L31.

\bibitem[\protect\astroncite{{Schwab}}{2018}]{schwab2018}
{Schwab}, J. 2018, \mnras, 476, 5303.

\bibitem[\protect\astroncite{{Shapiro}}{1964}]{shapiro1964}
{Shapiro}, I.~I. 1964, Physical Review Letters, 13, 789.

\bibitem[\protect\astroncite{{Sharma} \emph{et~al.}}{2016}]{sharmaetal2016}
{Sharma}, S., {Stello}, D., {Bland-Hawthorn}, J., {Huber}, D., \& {Bedding},
  T.~R. 2016, \apj, 822, 15.

\bibitem[\protect\astroncite{{Sharma} \emph{et~al.}}{2017}]{sharmaetal2017}
{Sharma}, S., {Stello}, D., {Huber}, D., {Bland-Hawthorn}, J., \& {Bedding},
  T.~R. 2017, \apj, 835, 163.

\bibitem[\protect\astroncite{{Shibahashi} \&
  {Kurtz}}{2012}]{shibahashi&kurtz2012}
{Shibahashi}, H. \& {Kurtz}, D.~W. 2012, \mnras, 422, 738.

\bibitem[\protect\astroncite{{Shibahashi} \&
  {Murphy}}{2018}]{shibahashi&murphy2018}
{Shibahashi}, H. \& {Murphy}, S.~J. 2018, ArXiv e-prints.

\bibitem[\protect\astroncite{{Shporer}}{2017}]{shporer2017}
{Shporer}, A. 2017, \pasp, 129, 072001.

\bibitem[\protect\astroncite{{Silvotti} \emph{et~al.}}{2007}]{silvottietal2007}
{Silvotti}, R., {Schuh}, S., {Janulis}, R., {Solheim}, J.-E., {Bernabei}, S.,
  \emph{et~al.} 2007, \nat, 449, 189.

\bibitem[\protect\astroncite{{Silvotti} \emph{et~al.}}{2018}]{silvottietal2018}
{Silvotti}, R., {Schuh}, S., {Kim}, S.-L., {Lutz}, R., {Reed}, M.,
  \emph{et~al.} 2018, \aap, 611, A85.

\bibitem[\protect\astroncite{{Stebbins}}{1911}]{stebbins1911}
{Stebbins}, J. 1911, \apj, 34, 105.

\bibitem[\protect\astroncite{{Stello} \emph{et~al.}}{2017}]{stelloetal2017}
{Stello}, D., {Zinn}, J., {Elsworth}, Y., {Garcia}, R.~A., {Kallinger}, T.,
  \emph{et~al.} 2017, \apj, 835, 83.

\bibitem[\protect\astroncite{{Sterken}}{2005}]{sterken2005c}
{Sterken}, C. 2005, In \emph{The Light-Time Effect in Astrophysics: Causes and
  cures of the O-C diagram}, edited by C.~{Sterken}, \emph{Astronomical Society
  of the Pacific Conference Series}, vol. 335, p.~3.

\bibitem[\protect\astroncite{{Sun} \& {Arras}}{2018}]{sun&arras2018}
{Sun}, M. \& {Arras}, P. 2018, \apj, 858, 14.

\bibitem[\protect\astroncite{{Tassoul} \&
  {Tassoul}}{1992}]{tassoul&tassoul1992a}
{Tassoul}, J.-L. \& {Tassoul}, M. 1992, \apj, 395, 259.

\bibitem[\protect\astroncite{{Telting} \emph{et~al.}}{2014}]{teltingetal2014b}
{Telting}, J.~H., {Baran}, A.~S., {Nemeth}, P., {{\O}stensen}, R.~H., {Kupfer},
  T., \emph{et~al.} 2014, \aap, 570, A129.

\bibitem[\protect\astroncite{{Telting} \emph{et~al.}}{2012}]{teltingetal2012}
{Telting}, J.~H., {{\O}stensen}, R.~H., {Baran}, A.~S., {Bloemen}, S., {Reed},
  M.~D., \emph{et~al.} 2012, \aap, 544, A1.

\bibitem[\protect\astroncite{{Theme{\ss}l}
  \emph{et~al.}}{2018}]{themessletal2018}
{Theme{\ss}l}, N., {Hekker}, S., {Southworth}, J., {Beck}, P.~G., {Pavlovski},
  K., \emph{et~al.} 2018, \mnras, 478, 4669.

\bibitem[\protect\astroncite{{Thompson} \emph{et~al.}}{2012}]{thompsonetal2012}
{Thompson}, S.~E., {Everett}, M., {Mullally}, F., {Barclay}, T., {Howell},
  S.~B., \emph{et~al.} 2012, \apj, 753, 86.

\bibitem[\protect\astroncite{{Tohline}}{2002}]{tohline2002}
{Tohline}, J.~E. 2002, \araa, 40, 349.

\bibitem[\protect\astroncite{{Torres} \emph{et~al.}}{2011}]{torresetal2011}
{Torres}, K.~B.~V., {Lampens}, P., {Fr{\'e}mat}, Y., {Hensberge}, H.,
  {Lebreton}, Y., \emph{et~al.} 2011, \aap, 525, A50.

\bibitem[\protect\astroncite{{Tremblay} \emph{et~al.}}{2016}]{tremblayetal2016}
{Tremblay}, P.-E., {Cummings}, J., {Kalirai}, J.~S., {G{\"a}nsicke}, B.~T.,
  {Gentile-Fusillo}, N., \emph{et~al.} 2016, \mnras, 461, 2100.

\bibitem[\protect\astroncite{{Tremblay} \emph{et~al.}}{2017}]{tremblayetal2017}
{Tremblay}, P.-E., {Gentile-Fusillo}, N., {Raddi}, R., {Jordan}, S., {Besson},
  C., \emph{et~al.} 2017, \mnras, 465, 2849.

\bibitem[\protect\astroncite{{Turcotte} \emph{et~al.}}{2000}]{turcotteetal2000}
{Turcotte}, S., {Richer}, J., {Michaud}, G., \& {Christensen-Dalsgaard}, J.
  2000, \aap, 360, 603.

\bibitem[\protect\astroncite{{Ulrich}}{1986}]{ulrich1986}
{Ulrich}, R.~K. 1986, \apjl, 306, L37.

\bibitem[\protect\astroncite{{Van Reeth}
  \emph{et~al.}}{2018}]{vanreethetal2018}
{Van Reeth}, T., {Mombarg}, J.~S.~G., {Mathis}, S., {Tkachenko}, A., {Fuller},
  J., \emph{et~al.} 2018, \aap, 618, A24.

\bibitem[\protect\astroncite{{Van Reeth}
  \emph{et~al.}}{2016}]{vanreethetal2016}
{Van Reeth}, T., {Tkachenko}, A., \& {Aerts}, C. 2016, \aap, 593, A120.

\bibitem[\protect\astroncite{{Van Reeth}
  \emph{et~al.}}{2015}]{vanreethetal2015b}
{Van Reeth}, T., {Tkachenko}, A., {Aerts}, C., {P{\'a}pics}, P.~I., {Triana},
  S.~A., \emph{et~al.} 2015, \apjs, 218, 27.

\bibitem[\protect\astroncite{{Viani} \emph{et~al.}}{2017}]{vianietal2017}
{Viani}, L.~S., {Basu}, S., {Chaplin}, W.~J., {Davies}, G.~R., \& {Elsworth},
  Y. 2017, \apj, 843, 11.

\bibitem[\protect\astroncite{{Webbink}}{1984}]{webbink1984}
{Webbink}, R.~F. 1984, \apj, 277, 355.

\bibitem[\protect\astroncite{{Welsh} \emph{et~al.}}{2011}]{welshetal2011}
{Welsh}, W.~F., {Orosz}, J.~A., {Aerts}, C., {Brown}, T.~M., {Brugamyer}, E.,
  \emph{et~al.} 2011, \apjs, 197, 4.

\bibitem[\protect\astroncite{{White} \emph{et~al.}}{2017}]{whiteetal2017a}
{White}, T.~R., {Benomar}, O., {Silva Aguirre}, V., {Ball}, W.~H., {Bedding},
  T.~R., \emph{et~al.} 2017, \aap, 601, A82.

\bibitem[\protect\astroncite{{Winget} \emph{et~al.}}{2015}]{wingetetal2015}
{Winget}, D.~E., {Hermes}, J.~J., {Mullally}, F., {Bell}, K.~J., {Montgomery},
  M.~H., \emph{et~al.} 2015, In \emph{19th European Workshop on White Dwarfs},
  edited by P.~{Dufour}, P.~{Bergeron}, \& G.~{Fontaine}, \emph{Astronomical
  Society of the Pacific Conference Series}, vol. 493, p. 285.

\bibitem[\protect\astroncite{{Winget} \& {Kepler}}{2008}]{winget&kepler2008}
{Winget}, D.~E. \& {Kepler}, S.~O. 2008, \araa, 46, 157.

\bibitem[\protect\astroncite{{Winget} \emph{et~al.}}{1990}]{wingetetal1990}
{Winget}, D.~E., {Nather}, R.~E., {Clemens}, J.~C., {Provencal}, J.,
  {Kleinman}, S.~J., \emph{et~al.} 1990, \apj, 357, 630.

\bibitem[\protect\astroncite{{Witte} \& {Savonije}}{2001}]{witte&savonije2001}
{Witte}, M.~G. \& {Savonije}, G.~J. 2001, \aap, 366, 840.

\bibitem[\protect\astroncite{{Yu} \emph{et~al.}}{2018}]{yuetal2018}
{Yu}, J., {Huber}, D., {Bedding}, T.~R., {Stello}, D., {Hon}, M., \emph{et~al.}
  2018, \apjs, 236, 42.

\bibitem[\protect\astroncite{{Zahn}}{1975}]{zahn1975}
{Zahn}, J.-P. 1975, \aap, 41, 329.

\bibitem[\protect\astroncite{{Zahn}}{1977}]{zahn1977}
{Zahn}, J.-P. 1977, \aap, 57, 383.

\bibitem[\protect\astroncite{{Zhao} \emph{et~al.}}{2012}]{zhaoetal2012}
{Zhao}, J.~K., {Oswalt}, T.~D., {Willson}, L.~A., {Wang}, Q., \& {Zhao}, G.
  2012, \apj, 746, 144.

\bibitem[\protect\astroncite{{Zong} \emph{et~al.}}{2018}]{zongetal2018}
{Zong}, W., {Charpinet}, S., {Fu}, J.-N., {Vauclair}, G., {Niu}, J.-S.,
  \emph{et~al.} 2018, \apj, 853, 98.

\end{thebibliography}

\end{document}